\documentstyle[12pt,twoside]{article}

\newcommand{\beqn}{\begin{eqnarray}}
\newcommand{\eeqn}{\end{eqnarray}}
\newcommand{\be}{\begin{equation}}
\newcommand{\ee}{\end{equation}}
\newcommand{\ba}{\begin{array}}
\newcommand{\ea}{\end{array}}
\newcommand{\q}{{\bf q}}
\newcommand{\R}{{\rm\bf R}}
\newcommand{\C}{{\rm\bf C}}
\newcommand{\pa}{\partial}
\newcommand{\re}{\ref}
\newcommand{\ci}{\cite}
\newcommand{\la}{\label}
\newcommand{\bfr}{\begin{flushright}}
\newcommand{\efr}{\end{flushright}}
\newcommand{\bfl}{\begin{flushleft}}
\newcommand{\efl}{\end{flushleft}}
\newcommand{\fr}{\frac}
\newcommand{\ov}{\overline}

\newcommand{\si}{\sigma}
\newcommand{\Si}{\Sigma}
\newcommand{\al}{\alpha}

\newcommand{\ds}{\displaystyle}
\newcommand{\ve}{\varepsilon}

\newcommand{\de}{\delta}
\newcommand{\De}{\Delta}
\newcommand{\na}{\nabla}
\newcommand{\ga}{\gamma}
\newcommand{\om}{\omega}
\newcommand{\Om}{\Omega}
\newcommand{\Lam}{\Lambda}
\newcommand{\lam}{\lambda}
\newcommand{\La}{\Lambda}

\newcommand{\br}{|\kern-.25em|\kern-.25em|}
\newcommand{\brr}{{|\kern-.15em|\kern-.15em|\kern-.15em}\,}

\begin{document}

\renewcommand{\theequation}{\thesection.\arabic{equation}}
\newtheorem{theorem}{Theorem}[section]
\renewcommand{\thetheorem}{\arabic{section}.\arabic{theorem}}
\newtheorem{definition}[theorem]{Definition}
\newtheorem{deflem}[theorem]{Definition and Lemma}
\newtheorem{lemma}[theorem]{Lemma}
\newtheorem{example}[theorem]{Example}
\newtheorem{remark}[theorem]{Remark}
\newtheorem{remarks}[theorem]{Remarks}
\newtheorem{cor}[theorem]{Corollary}
\newtheorem{pro}[theorem]{Proposition}
\mathsurround=2pt
\newcommand{\HH}{{\rm\bf H}}
\newcommand{\PV}{{\rm PV}}
\newcommand{\pr}{\prime}
\def\N{{\rm I\kern-.1567em N}}                              
\def\No{\N_0}                                   
\def\R{{\rm I\kern-.1567em R}}                              
\def\C{{\rm C\kern-4.7pt                                    
\vrule height 7.7pt width 0.4pt depth -0.5pt \phantom {.}}\,}
\def\Z{{\sf Z\kern-4.5pt Z}}                                
\def\n#1{\vert #1 \vert}                   
\def\nn#1{\Vert #1 \Vert}                    
\def\Re {{\rm Re\, }}                                       
\def\Im {{\rm Im\,}}                                        
\newcommand{\loc}{\scriptsize loc}
\newcommand{\tg}{\mathop{\rm tg}\nolimits}
\newcommand{\const}{\mathop{\rm const}\nolimits}
\newcommand{\sgn}{\mathop{\rm sgn}\nolimits}
\newcommand{\tr}{\mathop{\rm tr}\nolimits}
\newcommand{\supp}{\mathop{\rm supp}\nolimits}
\newcommand{\mod}{\mathop{\rm mod}\nolimits}
\newcommand{\ow}{\overrightarrow}

\begin{titlepage}
 \begin{center}
 {\Large\bf  On
 Two-Temperature Problem
\bigskip\\
 for Harmonic Crystals}\\
 \vspace{2cm}
{\large T.V.~Dudnikova}
\footnote{Supported partly by
research grants of DFG (436 RUS 113/615/0-1),
of RFBR (99-01-04012) and the START project
"Nonlinear  Schr\"odinger and Quantum Boltzmann Equations''
 (FWF Y 137-TEC) of N.J.~Mauser}\\
{\it  M.V.Keldysh Institute\\
of Applied Mathematics RAS\\
 Moscow 125047, Russia}\\
e-mail:~dudnik@elsite.ru, dudnik@mat.univie.ac.at
 \bigskip\\
 {\large A.I.~Komech}
\footnote{
 On leave Department of Mechanics and Mathematics,
Moscow State University, Moscow 119899, Russia.
Supported partly
by Max-Planck Institute for Mathematics in the Sciences (Leipzig).
}\\
{\it Instit\"ut f\"ur Mathematik\\
Wien A-1090, Austria}\\
 e-mail:~komech@mat.univie.ac.at
 \bigskip\\
 {\large N.J.~Mauser}\\
{\it Instit\"ut f\"ur Mathematik\\
Wien A-1090, Austria}\\
 e-mail:~mauser@courant.nyu.edu\\
 \end{center}
 \vspace{1cm}

 \begin{abstract}
We consider the dynamics of
a harmonic crystal in $d$ dimensions with $n$ components,
  $d,n \ge 1$.
The initial date  is a  random function
with finite mean density of the energy
which also satisfies  a Rosenblatt- or
Ibragimov-Linnik-type mixing condition.
The random function
converges to different space-homogeneous
processes as $x_d\to\pm\infty$, with the distributions  $\mu_\pm$.
We study the distribution $\mu_t$
 of the  solution at time $t\in\R$.
The main result is the convergence of $\mu_t$ to
 a Gaussian translation-invariant measure as $t\to\infty$.
The proof  is based on
the long time asymptotics
 of the Green function
 and on
  Bernstein's `room-corridor'  argument.
The application to the case of the Gibbs measures
$\mu_\pm=g_\pm$ with two different
temperatures $T_{\pm}$  is given.
Limiting mean energy current density
is $- (0,\dots,0,C(T_+\!\!-\!T_-))$
 with some positive constant $C>0$
what corresponds to Second Law.

 \end{abstract}
\end{titlepage}

\section{Introduction}
The paper concerns the problems related to the derivation
of Fourier's Law for harmonic crystals \ci{BLR}.
 We have started in  \cite{DKKS}-\cite{DKS1}
the analysis of the long time convergence
to the equilibrium distribution
for partial differential equations of
hyperbolic type in $\R^d$ and for a harmonic crystal.
Here we continue the analysis and prove
 Second Law  for the crystal: 
 the energy current is
 directed from high  temperature to low temperature.
Similar results have
 been established in \ci{BPT, SL}
for $d=1$, and we extend the results to all
$d\ge 1$.
The case $d>1$ appears very different from
$d=1$ because of
 much 
more complicated properties of oscillatory integrals. This is why we combine
here the methods from
\ci{BPT} with new ideas.
Namely, we develop our ``cutoff" strategy
from \ci{DKS1} which more carefully exploits the mixing condition
in Fourier space.
This approach allows us to cover all $d\ge 1$.

We assume that
the initial state $Y_0(x)$ of the crystal is a
random element of the Hilbert space ${\cal H}_\alpha$ of real sequences,
see Definition \ref{d1.1} below.
  The distribution of $Y_0(x)$ is a probability measure  $\mu_0$
of mean zero satisfying 
conditions {\bf S1}-{\bf S3} below.
In particular,
the distribution of $Y_0(x)$
converges to distinct translation-invariant
measures $\mu_\pm$ as $x_d\to\pm\infty$.  
Given $t\in\R$, denote by $\mu_t$ the probability measure
that gives the distribution of the
 solution $Y(x,t)$ to dynamical equations
with the random initial state $Y_0$.
 We study the
asymptotics of $\mu_t$ as $t\to\pm\infty$.

Our main result gives  the (weak)
convergence of the measures
$\mu_t$ on the Hilbert
space ${\cal H}_\al$
with $\al<-d/2$ to a limit measure $\mu_{\infty}$
\be\la{1.8i}
\mu_t \rightharpoondown
\mu_\infty,\,\,\,\, t\to \infty,
\ee
which is a translation-invariant Gaussian measure on
${\cal H}_\al$.
 A similar convergence result
holds for  $t\to-\infty$
since our system is time-reversible.
We construct generic examples of
harmonic crystals and
random initial data satisfying
all assumptions imposed.
The explicit formulas for the  covariance
 of  the measure $\mu_\infty$
 are  given in (\re{Qinfty})-(\re{qinfty-}).
We derive the expression for the limit
 mean energy current
$j_\infty$.
The ergodicity and mixing of the limit measures
$\mu_\infty$ follow by the same arguments as in \ci{DKS1}.
\medskip

We apply (\ref{1.8i}) to the case of the Gibbs measures
$\mu_\pm=g_{\pm}$ with two distinct   temperatures
 $T_\pm\ge0$
(we adjust the definition of the Gibbs measures $g_{\pm}$
in Section 3).
The  measures $g_\pm$ satisfy all our assumptions,
and  the weak
convergence $g_t\rightharpoondown g_\infty$
follows
from our results.
We apply formula 
for the limit energy current
$j_\infty\equiv
(j^1_\infty,\dots,j^d_\infty)$
to the case of Gibbs measures $g_\pm$
and deduce that 
$$
 j_\infty=- (0,\dots,0,C(T_+-T_-)),\,\,\,\,C>0.
$$
This corresponds to Second Law.

For $d=1$ similar problem  have been 
analyzed also in \ci{Na, RLL}. The authors considered 
the finite simple lattice of size $L$ with the viscosity,
in contact with  two heat baths at temperatures
$T_\pm$. The convergence of the covariance is proved
in the limit $t\to\infty$, and  then $L\to\infty$. 
The result is close to ours: the limit energy current is non
zero and $\sim\De T$ which corresponds to
the superconductivity \ci{BLR}.
However, the space decay of the limit
position-momentum covariance in \ci{Na} is exponential
which differs from
the power decay in our problem and in \ci{BPT}
(see Remark \ref{rema} iii)).
\medskip

For case $d\ge 1$ the convergence (\ref{1.8i}) has been obtained
for the first time in
\cite{LL} for initial measures which are absolutely
continuous with respect to the canonical Gaussian measure.
We cover more general class of
 initial measures with the mixing condition and do not assume the absolute
continuity.
For the first time
the mixing condition has been introduced by R.Dobrushin and
Yu.Suhov for the ideal gas \cite{DS}. The condition
substitutes (quasi-) ergodic hypothesis in the proof of the convergence
to the equilibrium distribution, and plays the key role in our
 Bernstein-type approach.
Developing this approach,
we have proved the convergence for the wave and Klein-Gordon equations
with translation-invariant initial measures
 \cite{DKKS,DKRS, K3}.
In \cite{DKS} we have extended the results to the
wave equation with
the two-temperature
initial measures.
The present paper develops our previous results
\ci{DKSab1, DKS1}, where the harmonic
crystal has been considered 
for all $d$ in the case of translation invariant initial measure.
Here we extend the results to the two-temperature
initial measures.
\medskip

We outline our main result and strategy
of proof.
Consider a discrete subgroup
$\Gamma$ of  $\R^d$, which is
isomorphic to $\Z^d$.
We may assume $\Gamma=\Z^d$ after
a suitable change of coordinates.
A {\it lattice} in $\R^d$
is the set of the points
of the form
$\ov r_\lam(x)=x+\xi_\lam$,
where $x\in\Z^d$, $\xi_\lam \in\R^d$,
 $\lam=
1,\dots,\Lam$.
The points of the lattice represent the
equilibrium positions of the atoms (molecules, ions,...)
of the crystal. Denote by $r_\lam(x,t)$ the positions of the atoms
in the dynamics. Then the dynamics of the
displacements $r_\lam(x,t)-\ov r_\lam(x)$ is governed by the equations of type
  \beqn\la{1.1'}
\left\{
\ba{l}
\ddot u(x,t)  =  -\sum\limits_{y\in\Z^d}
  V(x-y) u(y,t),~~
 x \in\Z^d,\\
u|_{t=0} = u_{0}(x),~~\dot u|_{t=0} = v_{0}(x).
\ea
\right.
\eeqn
Here
 $u(x,t)=(u_1(x,t),\dots,u_n(x,t)),
u_0=(u_{01},\dots,u_{0n}),
v_0=(v_{01},\dots,v_{0n})\in\R^n$,
$n=\Lambda d$;
$ V(x)$ is the real interaction 
(or force) matrix,
$ \Big(V_{kl}(x)\Big)$,
$k,l=1,...,n$.
Similar equations  were
 considered  in \ci{BPT, DKSab1, DKS1, LL, SL}.
Below we consider the system of type
(\re{1.1'}) with an arbitrary $n=1,2,...$.

Denote
$Y(t)=(Y^0(t),Y^1(t))\equiv
(u(\cdot,t),\dot u(\cdot,t))$, $Y_0=(Y^0_0,Y^1_0)\equiv
(u_0(\cdot),v_0(\cdot))$.
Then (\ref{1.1'}) takes the form of an evolution equation
\be\la{CP}
\dot Y(t)={\cal A}Y(t),\,\,\,t\in\R;\,\,\,\,Y(0)=Y_0.
\ee
Formally, this is the  Hamiltonian system since 
\be\la{A}   
{\cal A}Y=  
J 
\left(   
 \begin{array}{cc}   
{\cal V}& 0\\   
0 & 1   
\end{array}\right)Y 
=  
J 
\na H(Y),\,\,\,\,\,\,\,\,\,J= 
\left(   
 \begin{array}{cc}   
0 & 1\\   
- 1 & 0   
\end{array}\right). 
\ee   
Here ${\cal V}$ is a convolution operator with 
the matrix kernel $V$ 
and $H$ is  
the Hamiltonian functional  
\be\la{H}
H(Y):=
\frac{1}{2} \langle v,v\rangle
+\frac{1}{2}
\langle  {\cal V}u, u\rangle,
\quad Y=(u,v),
\ee
where the kinetic energy is given by
$\ds\frac{1}{2} \langle v,v\rangle=
\frac{1}{2} \sum_{x\in\Z^d}|v(x)|^2$
and the potential energy by
$\ds\frac{1}{2}\langle{\cal V}  u,u\rangle
=\frac{1}{2}\sum_{x,y\in\Z^d}
\Big(  V(x-y)u(y), u(x)\Big)$,
 $\Big(\cdot,\cdot\Big)$
stands for the real scalar product in the Euclidean space
$\R^n$.

We assume that
the initial
correlation functions
\be\la{1.9'}
Q^{ij}_0(x,y):= E\Big(Y_0^i(x)\otimes Y_0^j(y) \Big),\,\,\,x,y\in\Z^d,
\ee
have the  form
\be\la{3}
Q_0^{ij}(x,y)
=q_0^{ij}(\ov x-\ov y,x_d,y_d),\,\,\,i,j=0,1.
\ee
Here
$x= (x_1,\dots,x_d)\equiv(\ov x,x_d),$
$ y= (y_1,\dots,y_d)\equiv(\ov y,y_d)\in\Z^d$.
Moreover, we assume that
\be\la{3a}
\lim_{y_d\to \pm\infty}
q_0^{ij}(\ov z,y_d+z_d,y_d)=
q_\pm^{ij}(z), \,\,\, z=(\ov z,z_d)\in\Z^d.
\ee
Here $q_{\pm}^{ij}(z)$ are
 the correlation functions of some
translation-invariant measures $\mu_{\pm}$
with zero mean value  in ${\cal H}_\al$.
The measure  $\mu_0$ is not translation-invariant
if $q_-^{ij}\not=q_+^{ij}$.

Next,
we assume that
the
initial mean ``energy'' density  is uniformly bounded:
\be\la{med}
e_0(x):= E [
\vert  u_0(x)\vert^2
 + \vert v_0(x)\vert^2]
={\rm tr}\,Q_0^{00}(x,x)+{\rm tr}\,Q_0^{11}(x,x)
\le e_0
<\infty,\quad
 x\in\Z^d.
\ee
Finally, it is  assumed
 that the measure $\mu_0$ satisfies
a mixing
condition
of a Rosenblatt- or Ibragimov-Linnik type, which means that
\be\la{mix}
Y_0(x)\,\,\,\,   and \, \, \,\,Y_0(y)
\,\,\,\,  are\,\,\,\, asymptotically\,\,\,\, independent\,\, \,\,
 as \,\, \,\,
|x-y|\to\infty.
\ee
To  prove the convergence (\re{1.8i})
we  follow the strategy of
\ci{DKKS}--\ci{DKS1}.
There are  three steps:\\
{\bf I.} The family of measures
 $\mu_t$, $t\geq 0$, is weakly
compact in ${\cal H}_\al$, $\al<-d/2$.\\
{\bf II.} The correlation functions
converge to a limit,
 \be\la{corf}
Q^{ij}_t(x,y)
\equiv\ds\int \Big(Y^i(x)\otimes Y^j(y)\Big)\, \mu_t(dY)
\to Q^{ij}_\infty(x ,y),\,\,\,\,t\to\infty.
\ee
{\bf III.}
The characteristic functionals converge
to a Gaussian one,
\be\la{2.6i}
 \hat\mu_t(\Psi ):
 =   \int e^{i\langle Y,\Psi\rangle}\mu_t(dY)
\rightarrow \exp\{-\fr{1}{2}{\cal Q}_\infty (\Psi ,\Psi)\},\,\,\,
t\to\infty.
 \ee
Here $\Psi=(\Psi^0,\Psi^1)
\in{\cal D}:=D\oplus D$,
$D:=C_{0}(\Z^d)\otimes \R^n$,
where $C_{0}(\Z^d)$ denotes a space of real
sequences with
finite support,
$\langle Y,\Psi \rangle
=\sum\limits_{i=0,1}\sum\limits_{x\in\Z^d}\Big( Y^i(x),
\Psi^i(x)\Big)$,
and
${\cal Q}_\infty$ is
the
quadratic form
with the matrix  kernel
$(Q^{ij}_\infty(x,y))_{i,j=0,1}$,
\be\la{qpp}
{\cal Q}_\infty (\Psi, {\Psi})=\sum\limits_{i,j=0,1}~
\sum\limits_{x,y\in\Z^d}
\Big(Q_{\infty}^{ij}(x,y),\Psi^i(x)\otimes\Psi^j(y)\Big).
\ee
Below the brackets $\langle\cdot,\cdot\rangle$
denote also the Hermitian scalar product
in the Hilbert spaces
$L^2(T^d)\otimes\R^n$ or its 
 different extensions.

For the proof of  {\bf I} -- {\bf III} 
we develop our {\it cutting strategy} from \ci{DKS1}
combined with some techniques from  \ci{BPT}.
Each method  \ci{DKS1}
and \ci{BPT} separately is not
sufficient since here the measures $\mu_t$
are not translation-invariant and we consider
all $d\ge 1$.
To prove {\bf II} 
we  split $Q_t^{ij}(x,y)$
into  even, odd components and the remainder as in
 \ci{BPT}.
The even component corresponds to the 
translation-invariant
initial measure and is analyzed 
 by the method of \ci{DKS1} for all $d\ge 1$.
On the other hand, the odd component 
is missing in  \ci{DKS1} and  it requires a novel
idea since its  Fourier transform   contains
 the Cauchy Principal Value
which is  more  singular than 
 measures corresponding to
the even component.  
The singularity was studied  
in \ci{BPT} for the case $d=1$.
However,   similar detailed analysis
for $d\ge 1$ seems to be impossible
due to  the bifurcations of the critical points.

Let us outline our method.
We rewrite (\ref{corf}) in the equivalent form
\be\la{corfpsi}
{\cal Q}_t(\Psi,\Psi)\to 
{\cal Q}_\infty(\Psi,\Psi),\,\,\,t\to\infty,
\ee
for $\Psi\in{\cal D}^0$:
by definition of ${\cal D}^0$,
the  Fourier transform $\hat \Psi(\theta)$
vanishes in a neighborhood of a ``critical set''
${\cal C}\subset T^d$.
The set ${\cal C}$ includes
 all points $\theta\in T^d$
with a degenerate Hessian
 of $\om_k(\theta)$,
where $\om^2_k(\theta)$
are the eigenvalues of the matrix
$\hat V(\theta)= \sum_{z\in\Z^d}
e^{iz\theta}V(z)$.
Also the set ${\cal C}$
includes the points $\theta\in T^d$
either with  $\om_k(\theta)=0$, or
$\nabla_{\theta_d}\om_k(\theta)=0$ or
with non-smooth  $\om_k(\theta)$.
The cutting  of the critical set ${\cal C}$
 is possible by two key observations:
i) mes${\cal C}=0$ and ii) 
the correlation quadratic form is continuous in $l^2$
due to the mixing condition.
The continuity
follows from the space decay of correlation
functions by well-known Shur's lemma.
The systematic application of the Shur lemma allows
 us
 to extend  (\ref{corfpsi})
 from $\Psi\in{\cal D}^0$ to all $\Psi\in{\cal D}$
by condition {\bf E6}.

Similarly, we first
 prove the property {\bf III}
for $\Psi\in{\cal D}^0$
and then extend it to all $\Psi\in{\cal D}$.
For $\Psi\in{\cal D}^0$ we use a variant
of the S.N.~Bernstein `room-corridor' technique
(cf. \ci{BPT} for $d=1$).
We develop our variant
of the S.N.~Bernstein technique which we have
 introduced in
\ci{DKKS}--\ci{DKS}, \ci{K3}
in the context of the Klein-Gordon and wave equations
and in \ci{DKS1} for the harmonic crystal with
 $d\ge 1$
in the case of translation-invariant
 initial measures.
For
$\Psi\in{\cal D}^0$ we have
$\hat \mu_t(\Psi)=E\exp{(i\langle Y(t),\Psi\rangle)}$.
We rewrite,
$\langle Y(t),\Psi\rangle=
\langle Y(0),\Phi(\cdot,t)\rangle$,
where $\Phi(x,t)$
 and can be represented as an oscillatory integral.
For $\Phi(x,t)$ we get the uniform bounds (\ref{bphi}),
(\ref{conp}).
These bounds follow by the stationary phase method
because $\Phi(x,0)=\Psi(x)\in{\cal D}^0$,
and hence, $\hat\Psi(\theta)$
vanishes in all points $\theta\in{\cal C}$
with degenerate Hessian of the phase function.
The bounds roughly speaking 
imply   the following  representation:
\be\la{CLTi}
\langle Y(\cdot,t),\Psi\rangle
\sim\fr{\ds\sum\limits_{y\in B_t}
Y_0(y)}
{\sqrt{|B_t|} }\,\,,\,\,\,\,\,\,t\to\infty,
\ee
where  $B_t$ stands for the ball
$\{y\in\Z^d:|y|\le ct\}$ and
$|B_t|$ is its volume. 
 Now (\ref{2.6i})
follows from (\ref{CLTi}) by the Lindeberg
Central Limit Theorem 
since
  $Y_0(y_1)$,  $Y_0(y_2)$ are almost independent
for large $|y_1-y_2|$
by mixing condition (\ref{mix}).

Let us comment on our conditions concerning 
the interaction matrix $V(x)$. 
We assume the conditions   
{\bf E1-E4} below which  
in a similar form appear also in \ci{BPT,LL}.   
{\bf E1} means the exponential space-decay of the 
interaction in the crystal.
{\bf E2} resp. {\bf E3}  means that the potential energy
 is real resp. nonnegative.
{\bf E4}  eliminates the discrete
 part of the spectrum  
and ensures   
 that  mes${\cal C}=0$. 
We also introduce a new simple condition {\bf E5}  
for the case $n>1$
which eliminates the {\it discrete} part of the spectrum
for the covariance dynamics.
It can be considerably weakened 
to the condition 
{\bf E5'} from Remark \ref{remi-iii} $iii)$.  
For example, the condition {\bf E5'} 
holds for the canonical Gaussian measures  
which are considered in \ci{LL}.
The conditions {\bf E4}, {\bf E5}  hold for almost
all functions $V(x)$ satisfying {\bf E1-E3}
as shown in \ci{DKS1}.
Furthermore, we do not require that 
$\om_k(\theta)\ne 0$, as in \ci{BPT}:
note that $\om(0)=0$ for  the elastic lattice   
(\ref{omega})  in the case $m=0$.   
Our results  hold whenever 
mes$\{\theta\in T^d:\om_k(\theta)\!=\!0\}\!=\!0$.  
To cover this case we  impose  the  
 condition {\bf E6} which
is similar to the condition iii)
from  \ci[p.171]{LL}. {\bf E6} 
 holds for the elastic lattice  
(\ref{omega}) if either $n\ge 3$ or $m > 0$.  

The main result of the paper is stated in Section 2
(see Theorem A).
Section 3 concerns  the application to Gibbs
measures.
In Section 4 we give bounds for the initial covariance.
The compactness (Property {\bf I})
 is established in Section 5,
convergence (\re{corf})   in Sections 6-7,
and convergence (\re{2.6i}) in Section 8.
In Section 9 we  check the Lindeberg condition
 for convergence to a Gaussian limit.
Appendix is concerned with a dynamics and
covariance in Fourier space.

\setcounter{equation}{0}
 \section{Main results}
Let us describe our results more precisely.
\subsection{Dynamics}
We assume that the initial date $Y_0$
belongs to the phase space ${\cal H}_\al$,
 $\al\in\R^1$,
defined below.
 \begin{definition}                 \la{d1.1}
  $ {\cal H}_\al$
 is the Hilbert space
of pairs $Y\equiv(u(x),v(x))$  of
 $\R^n$-valued functions  of $x\in\Z^d$
 endowed  with  the  norm
 \beqn                              \la{1.5}
 \Vert Y\Vert^2_{\al}
 \equiv
 \sum\limits_{x\in\Z^d}\Big(
\vert u(x)\vert^2
 +  \vert v(x)\vert^2\Big)(1+|x|^2)^{\al}
  <\infty.
 \eeqn
  \end{definition}
We impose the following conditions {\bf E1}-{\bf E6}
 on the matrix
$V$.
\medskip\\
{\bf E1}
There exist constants $C,\alpha>0$ such that
$
|V_{kl}(z)|\le C e^{-\alpha|z|},\,\,\,\,
 k,l \in \ov n:=\{1,...,n\},\,\,\,\,z\in \Z^d.
$
\smallskip\\
Let us denote  by
$
\hat V(\theta):=
\Big(\hat V_{kl}(\theta)
\Big)_{k,\,l\in\ov n},
$
where
$
\hat V_{kl}(\theta)\equiv
\sum\limits_{z\in\Z^d}V_{kl}(z)e^{iz\theta },$
$\theta \in T^d,$
and  $T^d$ denotes the $d$-torus
$\R^d/{2\pi \Z^d}$.
\medskip\\
{\bf E2} $ V$ is real and
symmetric, i.e.
$V_{lk}(-z)=V_{kl}(z)\in \R$, $k,l \in \ov n$,
$z\in \Z^d$.
\smallskip\\
The condition implies that $\hat V(\theta)$ is
 a real-analytic
 Hermitian matrix-function in $\theta\in T^d$.
\medskip\\
{\bf E3}
The matrix
$\hat V(\theta )$
is  non-negative definite for each
 $\theta \in T^d.$
\medskip\\
The condition means that the Eqn  (\re{1.1'}) is
hyperbolic like
wave and Klein-Gordon Eqns considered in \ci{DKKS,DKRS}.
Let us define the Hermitian  non-negative definite
 matrix
\be\la{Omega}
\Omega(\theta ):=\big(\hat V(\theta )\big)^{1/2}\ge 0
\ee
with  the eigenvalues  $\om_k(\theta)\ge 0$,
$k\in\ov n$, which
 are called  dispersion relations.
For each $\theta\in T^d$ the Hermitian matrix
$\Om(\theta)$  has the diagonal form
 in the  basis  of the orthogonal
eigenvectors $\{e_k(\theta):k\in\ov n \}$:
\be\la{diag}
\Om(\theta)=B(\theta)
\left(\ba{ccc}\om_1(\theta)&\ldots& 0\\
0&\ddots&0\\
0&\ldots&\om_n(\theta)
\ea
\right)B^*(\theta),
\ee
where $B(\theta)$ is a unitary matrix
and $B^*(\theta)$ denotes its adjoint.
It is well known that the functions
$\om_k(\theta)$ and  $B(\theta)$ are real-analytic
outside the set of the `crossing' points $\theta_*$:
$\om_k(\theta_*)=\om_l(\theta_*)$ for some  $l\ne k$.
However,
generally  the functions  are not smooth
at the crossing points if $\om_k(\theta)\not\equiv\om_l(\theta)$.
Therefore, we need the following lemma which is proved in
\ci[Appendix]{DKS1}.
\begin{lemma}\la{lc*}
Let the conditions {\bf E1}, {\bf E2} hold. Then
there exists
a closed subset ${\cal C}_*\subset T^d$ such that
i) the Lebesgue measure of ${\cal C}_*$ is zero:
\be\la{c*}
{\rm mes}\,{\cal C}_*=0.
 \ee
ii) For any point $\Theta\in T^d\setminus{\cal C}_*$
there exists a neighborhood
${\cal O}(\Theta)$ such that each dispersion relation
$\om_k(\theta)$ and the matrix $B(\theta)$
can be chosen
as
the real-analytic functions in
${\cal O}(\Theta)$.
\\
iii) It is possible to enumerate the eigenvalues
$\om_k(\theta)$ so that
we have  in the whole open set
 $T^d\setminus{\cal C}_*$:
\beqn
&&\om_1(\theta)\!\equiv\!\dots\!\equiv\!\om_{r_1}(\theta),\,\,\,
\om_{r_1+1}(\theta)\!\equiv\!\dots\equiv\!\om_{r_2}
(\theta),\,\,\,
\dots \,\,\,, \om_{r_s+1}(\theta)\!\equiv\!\dots\!\equiv\!
\om_{n}(\theta)
,\la{enum}\\
&&\om_{r_\si}(\theta)\!\not\equiv\!
\om_{r_\nu}(\theta)\,\,\,\,{\rm if}\,\,\,\,\si\ne\nu,
\,\,\,1\le r_\si,r_\nu\le r_{s+1}:=n.
\la{enum'}
\eeqn
iv) Let us
define $\Pi_\si(\theta)$
for $\theta\in T^d\setminus{\cal C}_*$
as the
orthogonal projection
$\Pi_\si(\theta):\R^n\to E_\si(\theta)$ onto the eigenspace
 $E_\si(\theta)\subset\R^n$
generated by the eigenvectors
$e_k(\theta)$, $k\in(r_{\si-1}, r_\si]$. Then
$\Pi_\si(\theta)$, $\si=1,...,s+1$, is
a
real-analytic function of
$\theta\in T^d\setminus{\cal C}_*$.
\end{lemma}
Below we suggest that $\om_k(\theta)$ denote the local
real-analytic
functions from Lemma \re{lc*} $ii)$.
Our next condition is the following:
\medskip\\
{\bf E4}
$D_k(\theta)\not\equiv 0$,
$\forall k\in\ov n $, where
$D_k(\theta):=
\det\Big(
\ds\frac{\pa^2\om_k(\theta)}{\pa \theta_i\pa \theta_j}
\Big)_{i,j=1}^{n}$, $\theta\in T^d\setminus{\cal C}_* $.
\medskip\\
Let us denote ${\cal C}_0:=\{\theta\in T^d:\det \hat V(\theta)=0\}$
and ${\cal C}_k:=\{\theta\in T^d\setminus
{\cal C}_*:\,D_k(\theta)=0\}$, $k=1,\dots,n$.
The following lemma
has also
been proved in \ci[Appendix]{DKS1}.
\begin{lemma}\la{lc}
Let the conditions {\bf E1}--{\bf E4} hold. Then
${\rm mes}~{\cal C}_k=0$, $k=0,1,...,n.$
\end{lemma}
Our last conditions on $V$ are the following:
\medskip\\
{\bf E5}
For each $k\ne l$  the identity 
$\om_k(\theta)-\om_l(\theta)\!\equiv\!{\rm const}_-$, $\theta\in T^d$   
does not hold  with const$_-\ne 0$, and
the identity 
$\om_k(\theta)+\om_l(\theta)\!\equiv\!{\rm const}_+$ 
does not hold  with const$_+\ne 0$.   
\medskip\\
{\bf E6} $\Vert \hat V^{-1}(\theta)\Vert\in L^1(T^d)$
in the case when ${\cal C}_0\ne\emptyset$.
\smallskip\\
This condition holds if ${\cal C}_0=\emptyset$.
\medskip

The following Proposition \re{p1.1} is proved in
\ci[p.150]{LL}, \ci[p.128]{BPT} (see also Appendix).
 \begin{pro}    \la{p1.1}
Let {\bf E1} and {\bf E2} hold,
and $\al\in\R$.
Then \\
i)
for any  $Y_0 \in {\cal H}_\al$
 there exists  a unique solution
$Y(t)\in C(\R, {\cal H}_\al)$
 to the Cauchy problem (\re{CP}).
\\
ii) The operator $U(t):Y_0\mapsto Y(t)$ is continuous in ${\cal H}_\al$.
\end{pro}

\subsection{Convergence to statistical
equilibrium}

Let $(\Om,\Si,P)$ be a probability space
with  expectation $E$
and ${\cal B}({\cal H}_\al)$ denote the Borel $\si$-algebra
in ${\cal H}_\al$.
We assume that $Y_0=Y_0(\om,\cdot)$ in (\re{CP})
is a measurable
random function
with values in $({\cal H}_\al,\,
{\cal B}({\cal H}_\al))$.
In other words,
for each $x\in \Z^d$ the map
$\om\mapsto Y_0(\om,x)$ is a measurable
 map
$\Om\to\R^{2n}$ with respect to the
(completed)
$\si$-algebras
$\Si$ and ${\cal B}(\R^{2n})$.
Then
$Y(t)=U(t) Y_0$ is again a measurable  random function
with values in
$({\cal H}_\al,{\cal B}({\cal H}_\al))$ owing  to Proposition \re{p1.1}.
We denote by $\mu_0(dY_0)$ a Borel probability measure
on ${\cal H}_\al$
giving
the distribution of the  $Y_0$.
Without loss of generality,
 we assume $(\Om,\Si,P)=
({\cal H}_\al,{\cal B}({\cal H}_\al),\mu_0)$
and $Y_0(\om,x)=\om(x)$ for
$\mu_0(d\om)$-almost all
$\om\in{\cal H}_\al$ and each $x\in  \Z^d$.

\begin{definition}
$\mu_t$ is a Borel probability measure in
${\cal H}_\al$
which gives
the distribution of $Y(t)$:
\begin{eqnarray}\la{1.6}
\mu_t(B) = \mu_0(U(-t)B),\,\,\,\,
\forall B\in {\cal B}({\cal H}_\al),
\,\,\,   t\in \R.
\eeqn
\end{definition}

Our main goal is to derive
 the convergence of the measures $\mu_t$ as $t\rightarrow \infty $.
We establish the weak convergence of  $\mu_t$
in the Hilbert spaces ${\cal H}_\al$
with $\al<-d/2$:
\be\la{1.8}
\mu_t\,\buildrel {\hspace{2mm}{\cal H}_\al}\over
{- \hspace{-2mm} \rightharpoondown }
\, \mu_\infty
\quad{\rm as}\quad t\to \infty,
\ee
where $\mu_\infty$ is a limit measure on the space
${\cal H}_\al$, $\al<-d/2$.
This means the convergence
 \be\la{1.8'}
 \int f(Y)\mu_t(dY)\rightarrow
 \int f(Y)\mu_\infty(dY),\quad t\to \infty,
 \ee
 for any bounded continuous functional $f$
 on ${\cal H}_\al$.

\begin{definition}
The correlation functions of the  measure $\mu_t$ are
 defined by
\be\la{qd}
Q_t^{ij}(x,y)= E \Big(Y^i(x,t)\otimes
 Y^j(y,t)\Big),\,\,\,i,j= 0,1,\,\,\,\,x,y\in\Z^d,
\ee
if the expectations in the RHS are finite.
Here $Y^i(x,t)$ are the components of the random solution
$Y(t)=(Y^0(\cdot,t),Y^1(\cdot,t))$.
\end{definition}
For a probability  measure $\mu$ on  ${\cal H}_\al$
we denote by $\hat\mu$
the characteristic functional (Fourier transform)
$$
\hat \mu(\Psi )  \equiv  \int\exp(i\langle Y,\Psi \rangle )\,\mu(dY),\,\,\,
 \Psi\in  {\cal D}.
$$
A measure $\mu$ is called Gaussian (of zero mean) if
its characteristic functional has the form
$$
\ds\hat { \mu} (\Psi ) =  \ds \exp\{-\fr12
 {\cal Q}(\Psi , \Psi )\},\,\,\,\Psi \in {\cal D},
$$
where ${\cal Q}$ is a real nonnegative quadratic form in ${\cal D}$.
A measure $\mu$ is called
translation-invariant if
$\mu(T_h B)= \mu(B)$, $B\in{\cal B}({\cal H}_\alpha)$,
 $h\in\Z^d$,
where $T_h Y(x)= Y(x-h)$, $x\in\Z^d$.

\subsection{Mixing condition}
Let $O(r)$ denote the set of all pairs of 
 subsets
${\cal A},\>{\cal B}\subset \Z^d$ at distance
dist$({\cal A},\,{\cal B})\geq r$ and let $\sigma ({\cal A})$
be a  $\sigma $-algebra  in ${\cal H}_\al$ generated by 
 $Y(x)$ with  $x \in {\cal A}$.
Define the
Ibragimov-Linnik mixing coefficient
of a probability  measure  $\mu_0$ on ${\cal H}_\al$
by (cf \ci[Definition 17.2.2]{IL})
\be\la{ilc}
\varphi(r)\equiv
\sup_{({\cal A},{\cal B})\in O(r)} \sup_{
\ba{c} A\in\si({\cal A}),B\in\si({\cal B})\\ \mu_0(B)>0\ea}
\fr{| \mu_0(A\cap B) - \mu_0(A)\mu_0(B)|}{ \mu_0(B)}.
\ee
\begin{definition}
 The measure $\mu_0$ satisfies strong, uniform
Ibragimov-Linnik mixing condition if
\be\la{1.11}
\varphi(r)\to 0\quad{\rm as}\quad r\to\infty.
\ee
\end{definition}
Below, we  specify the rate of decay of $\varphi$
(see condition {\bf S3}).

\subsection{Statistical conditions and results}

We assume that the initial measure $\mu_0$
satisfies
 the following conditions {\bf S0}-{\bf S3}:
\medskip\\
{\bf S0}  $\mu_0$ has  zero expectation value,
$EY_0(x)= 0$, $x\in\Z^d.$\\
{\bf S1} $\mu_0$ has correlation functions
of   the form (\re{3}) with condition
 (\re{3a}).\\
{\bf S2}
$\mu_0$ has a finite mean energy density, i.e.
Eqn  (\ref{med})
holds.\\
{\bf S3}  $\mu_0$ satisfies the  strong uniform
 Ibragimov-Linnik mixing condition with
\be\la{1.12}
 \ov\varphi\equiv\ds
\int\limits_0^{+\infty} r^{d-1}\varphi^{1/2}(r)dr <\infty.
 \ee
Introduce
 the correlation matrix of the limit
measure $\mu_\infty$.
It is translation-invariant
\be\la{Qinfty}
Q_\infty(x,y)=
 \Bigl(Q_\infty^{ij}(x, y)\Bigr)_{i,j= 0,1} =
 \Bigl(q_\infty ^{ij}(x-y)\Bigr)_{i,j= 0,1}\,\,.
\ee
In the Fourier transform we have
locally outside the critical set ${\cal C}_*$
(see Lemma \ref{lc*})
\be\la{qinfty}
\hat q_\infty^{ij}(\theta)=
B(\theta)M_{\infty}^{ij}(\theta)B^{*}(\theta),
\,\,\,\,i,j= 0,1,
\ee
where $B(\theta)$ is the smooth unitary matrix from
Lemma \ref{lc*}, {\it ii)}
and $M^{ij}_\infty(\theta)$ is $n\times n$-matrix
with the smooth entries
\beqn\la{Piinfty}
M^{ij}_\infty(\theta)_{kl}=\chi_{kl}
\left[
\Big(
B^{*}
(\theta)(M_0^+)^{ij}(\theta)
B(\theta)
\Big)_{kl}+
i\sgn\Big(\frac{\pa\om_k}{\pa\theta_d}(\theta)\Big)\Big(
B^{*}
(\theta)(M_0^-)^{ij}(\theta)
B(\theta)
\Big)_{kl}
\right].\,\,\,\,
\eeqn
Here  we set (see (\re{enum}))
\be\la{chi}
\chi_{kl}=
\left\{
\ba{rl}
1 &{\rm if} \,\,\,\, k,l\in(r_{\si-1}, r_{\si}],\,\,\,
\si=1,...,s+1, \\
 0& {\rm otherwise}
\ea
\right.
\ee
with $r_0:=0$, $r_{s+1}:=n$,
and
\beqn
M_0^+(\theta)&:=&\frac{1}{2}
\Big(\hat \q^+(\theta)+
\hat C(\theta)\hat \q^+(\theta)\hat C^*(\theta)\Big),\la{Pi+}\\
M_0^-(\theta)&:=&\frac{1}{2}
\Big(\hat C(\theta)\hat \q^-(\theta) -
 \hat \q^-(\theta)\hat C^*(\theta)\Big)\,,
\la{Pi-}
\eeqn
with
$\q^+:=\ds\frac{1}{2}(q_++q_-)$,
$\q^-:=\ds\frac{1}{2}(q_+-q_-)$
and
\be\la{C(theta)}
\hat C(\theta):=\left(
\ba{lcl}
0&\Om^{-1}(\theta)\\
-\Om(\theta)&0
\ea
\right),\,\,\,\,
\hat C^*(\theta):=\left(
\ba{lcl}
0&-\Om(\theta)\\
\Om^{-1}(\theta)&0
\ea
\right),
\ee
where $\hat C^*$ denotes a Hermitian
conjugate matrix to the matrix
$\hat C$.
The local representation 
(\re{qinfty}) can be expressed globally as 
the sum:
\be\la{QinftyFP}  
\hat q_\infty(\theta)=
\hat q^+_\infty(\theta)+\hat q^-_\infty(\theta),
\ee 
where 
\beqn\la{qinfty+}
(\hat q^+_\infty)^{ij}(\theta)&:=&
\sum_{\si=1}^{s+1}  \Pi_\si(\theta)(M_0^+)^{ij}(\theta)
\Pi_\si(\theta),  \\
\la{qinfty-}
(\hat q^-_\infty)^{ij}(\theta)&:=&
\sum_{\si=1}^{s+1}  \Pi_\si(\theta)
\,i\sgn\Big(\frac{\pa\om_{r_\sigma}}{\pa\theta_d}(\theta)\Big)
(M_0^-)^{ij}(\theta)
\Pi_\si(\theta),\,\,\,\,
\theta\in T^d\setminus{\cal C}_*,
\,\,\,\, i,j=0,1.\,\,\,\,\,\,\,\,\,\,
\eeqn
Here
$\Pi_\si(\theta)$ is the spectral projection 
introduced in Lemma \re{lc*} $iv)$.
\begin{remark}\la{rem} 
{\rm From Proposition \ref{l4.1}, ii)
and condition {\bf E6}  
(if ${\cal C}_0\not=0$) it follows that
$\Big( (M_0^\pm)^{ij}\Big)_{kl}\in L^1(T^d)$,  
$k,l\in\ov n$.  
Therefore,  
(\re{qinfty+}), (\re{qinfty-})
 and  (\re{c*}) 
imply that also $\Big((\hat q^\pm_\infty)^{ij}\Big)_{kl}\in L^1(T^d)$,  
$k,l\in\ov n$. 
}  
\end{remark} 
{\bf Theorem A}
{\it  Let $d,n\ge 1$, $\al<-d/2$
and assume that the conditions
{\bf E1--E5} and   {\bf S0--S3}
 hold.
If ${\cal C}_0\not=0$,
then we assume also that {\bf E6} holds.
 Then \\
i) the convergence (\re{1.8}) holds and  (\re{corf}) also holds.\\
ii) The limit measure
$ \mu_\infty $ is a Gaussian translation-invariant
measure on ${\cal H}_\al$.
\\
iii) The  characteristic functional of $ \mu_{\infty}$
is the Gaussian
$$
\ds\hat { \mu}_\infty (\Psi ) =
\exp\{-\fr{1}{2}  {\cal Q}_\infty  (\Psi ,\, \Psi)\},\,\,\,
\Psi \in {\cal D},
$$
where ${\cal Q}_\infty$
is the quadratic form defined in (\ref{qpp}).\\
iv) The measure $\mu_\infty$ is invariant, i.e.
$[U(t)]^*\mu_\infty=\mu_\infty$, $t\in\R$.
}
\begin{remarks}\la{remi-iii} 
{\rm  
{\it i)}
In the case $d=n=1$
we have $B(\theta)\equiv 1$, and
formulas (\ref{Qinfty})-(\ref{Pi-})
have been obtained in \ci[p.139]{BPT}.


{\it ii)}
 The {\it uniform} Rosenblatt mixing condition
\ci{Ros} also suffices, together with a higher
power $>2$  in the bound (\re{med}): there exists $\de >0$ such that
$$
E \Big(
\vert u_0(x)\vert^{2+\de}+\vert v_0(x)\vert^{2+\de}
\Big) \le C
<\infty.
$$
Then  (\re{1.12}) requires a modification:
$
\ds\int_0^{+\infty}\ds r^{d-1}\al^{p}(r)dr <\infty,$
where $p=\min(\de/(2+\de),  1/2),$
where $\al(r)$ is the  Rosenblatt
mixing coefficient  defined
as in  (\re{ilc}) but without $\mu_0(B)$ in the denominator.
Under these modifications, 
the statements of Theorem A and their
proofs remain essentially unchanged.

{\it iii)}
The arguments with condition {\bf E5}  
in Lemmas \ref{Qt1}, \ref{Qt3}
  demonstrate that the condition  
could be considerably weakened.  
Namely, it suffices to assume\\
{\bf E5'} If
for some $k\not=l$ we have
either
$\om_k(\theta)+\om_l(\theta)\equiv \const_+$
or
$\om_k(\theta)-\om_l(\theta)\equiv \const_-$
with $\const_\pm\not=0$,
then
either
$p_{kl}^{11}(\theta)-\om_k(\theta)\om_l(\theta)
p_{kl}^{00}(\theta)\equiv 0$,
$\om_k(\theta)p_{kl}^{01}(\theta)+
\om_l(\theta)p_{kl}^{10}(\theta)\equiv 0$
or
$p_{kl}^{11}(\theta)+\om_k(\theta)\om_l(\theta)
p_{kl}^{00}(\theta)\equiv 0$,
$\om_k(\theta)p_{kl}^{01}(\theta)-
\om_l(\theta)p_{kl}^{10}(\theta)\equiv 0$.
Here
\be\la{pij}
p_{kl}^{ij}(\theta):=
\Big(B^*(\theta)\hat q_\pm^{ij}(\theta)B(\theta)\Big)_{kl},\,\,\,\,
\theta\in T^d,\,\,\,\, k,l\in\ov n,\,\,\,\, i,j=0,1,
\ee
 $\hat q_\pm^{ij}(\theta)$
are Fourier transforms of covariance matrices
$q^{ij}_\pm(z)$.
}   
\end{remarks}

The assertions {\it i)-iii)} of Theorem~A follow  from Propositions  \re{l2.1} and \re{l2.2} below.
 \begin{pro}\la{l2.1}
  The family of the measures $\{\mu_t,\,t\in \R\}$
 is weakly compact in   ${\cal H}_\al$ with any
 $\al<-d/2$,
and the bounds 
$\sup\limits_{t\ge 0}
 E \Vert U(t)Y_0\Vert^2_\al <\infty
$ hold.
 \end{pro}
 \begin{pro}\la{l2.2}
  For every $\Psi\in {\cal D}$ the convergence  (\ref{2.6i})
holds.
 \end{pro}
Proposition  \ref{l2.1}
(Proposition \ref{l2.2})
 provides the existence
(resp. the uniqueness)
of the limit measure $\mu_\infty$.
They are proved
in Sections 5 and 7-8, respectively.

Theorem~A {\it iv)} follows from  
 (\re{1.8})   since the group $U(t)$ is continuous in  
${\cal H}_\al$ by Proposition~\re{p1.1} {\it ii)}. 
\subsection{Examples}
Let us give the examples of Eqn (\ref{1.1'})
and measures $\mu_0$ which
satisfy all our conditions
 {\bf E1-E6}
 and {\bf S0-S3}, respectively.
\subsubsection{Harmonic crystals}
All  conditions {\bf E1-E5}
hold for 1D crystal with $n=1$
considered in \ci{BPT}.
For any $d\ge 1$ and $n=1$ we  consider
the simple elastic lattice
 corresponding to the quadratic form   
\be\la{dKG}
\langle {\cal V}u,u \rangle=
\sum\limits_{x\in\Z^d}
(\sum\limits_{k=1}^{d}|u(x+e_k)-u(x)|^2+
m^2|u(x)|^2), \,\,\,m>0,
\ee
where $e_k=(\de_{k1},\dots, \de_{kd})$.
Then {\bf E1} holds and 
 $F_{x\to \theta}V=\om^2(\theta)$ with
\be\la{omega}
 \omega(\theta)=
\sqrt{\, 2(1-\cos\theta_1)+...+2(1-\cos\theta_d)+m^2}.
\ee
Hence, $V(x)$
satisfies {\bf E2-E4}
with ${\cal C}_*=\emptyset$.
In these examples the set 
${\cal C}_0$ is  
empty, hence  
the condition {\bf E6} is unnecessary.
Condition {\bf E5} holds trivially since $n=1$. 
Therefore, Theorem~A holds for (\ref{dKG})
if the initial measure $\mu_0$
satisfies  conditions {\bf S0-S3}.
In these examples
 $B(\theta)\equiv 1$,
and then,
$
 \hat q_\infty=M_{\infty}= M_0^+
+i\sgn\Big(\ds\pa_{\theta_d}\om\Big)M_0^-$,
where $M_0^\pm$ is defined by (\ref{Pi+})-(\ref{C(theta)})
with $\Omega(\theta)=\omega(\theta)$.
Then
we obtain the following explicit formulas
for $q_\infty$.
Denote by ${\cal E} (z)=F^{-1}_{\theta\to z}
\ds(\omega^{-2}(\theta))$ 
the fundamental solution for the
operator $-\De+m^2$   on    the lattice $\Z^d$,
i.e. $(-\De+m^2){\cal E}(x)=\delta_{0x}$ for  $x\in\Z^d$, and
$P(x)
=\ds-i
F^{-1}_{\theta\to x}\ds
\frac{\sgn(\sin \theta_d)}{\om(\theta)}$.
Then
\beqn
q^{00}_{\infty}&=&\ds\frac{1}{2}
\left[
(\q^+)^{00}+
 {\cal E}*(\q^+)^{11} +
P*\Big((\q^-)^{01}-(\q^-)^{10}\Big)\right],
\nonumber\\
q^{10}_{\infty}=\,\,\,-\,\,
q^{01}_{\infty}&=&
\ds\frac{1}{2}
\left[
(\q^+)^{10} -(\q^+)^{01}\,\,\,\,\,\,\,\,\,+
P*\Big((\q^-)^{11}+(-\De+m^2) (\q^-)^{00}\Big)\right],
\nonumber\\
q^{11}_{\infty}=(-\De+m^2) q^{00}_{\infty}&=&\ds\frac{1}{2}
\left[(\q^+)^{11} +
(-\De+m^2) \Big((\q^+)^{00} +
P*\Big((\q^-)^{01}-(\q^-)^{10}\Big)\Big)
\right],
\nonumber
\eeqn
where
$*$ stands  for the convolution of  functions.

\subsubsection{Gaussian measures}
We consider $n=1$
and construct
Gaussian initial measures $\mu_0$ satisfying {\bf S0}--{\bf S3}.
We will define
 $\mu_\pm$ in ${\cal H}_\al$
by the  correlation
functions $q_\pm^{ij}(x-y)$ which are zero for
 $i\not= j$, while for $i=0,1$,
\be\la{S04}
\hat q_\pm^{ii}(\theta):=F_{z\to\theta}
[ q_\pm^{ii}(z)]\in L^1(T^d),\,\,\,\,
\hat q_\pm^{ii}(\theta) \ge 0.
\ee
Then by the Minlos theorem, \ci{CFS}
there exist  Borel Gaussian measures
$\mu_{\pm}$ on  ${\cal H}_\al$,
$\al<-d/2$,
with the correlation functions
$q^{ij}_\pm(x-y)$,
because {\it formally} we have
$$
\int\!\Vert Y\Vert^2_\al\mu_\pm(dY)
=\sum\limits_{x\in\Z^d}\!(1\!+\!|x|^2)^\al
(\tr q^{00}_\pm(0)\!+\!\tr q_\pm^{11}(0))
=C(\al,d)\int\limits_{T^d}
\!\tr( \hat q^{00}_\pm(\theta)\!+\!\hat q_\pm^{11}(\theta))\,d\theta<\infty.
$$
The measures
$\mu_\pm$ satisfy {\bf S0}, {\bf S2}.
Let us take the functions $\zeta_{\pm}\in C(\Z)$ such that
\be\la{zeta}
\zeta_{\pm}(s)= \left\{ \ba{ll}
1,~~\mbox{for }~ \pm s>\,a,\\
0,~~\mbox{for }~ \pm s<-a.\ea\right.
\ee
Let us introduce
$(Y_-,Y_+)$ as
a unit random function in probability space
$({\cal H}_\al\times{\cal H}_\al, \mu_-\times\mu_+)$.
Then $Y_{\pm}$ are
 Gaussian independent vectors
in
${\cal H}_\al$.
Define
a ``two-temperature'' Borel probability measure
 $\mu_0$ as a
distribution of the random function
\be\la{rf}
Y_0(x)= \zeta_-(x_d)Y_-(x)+\zeta_+(x_d)Y_+(x).
\ee
Then correlation functions of  $\mu_0$
are
\be\la{q1}
Q_0^{ij}(x,y)=
q_-^{ij}(x-y)\zeta_-(x_d)\zeta_-(y_d)+
q_+^{ij}(x-y)
\zeta_+(x_d)\zeta_+(y_d),~~i,j= 0,1,
\ee
where $x= (x_1,\dots,x_d)$,
$y= (y_1,\dots,y_d)\in \Z^d$,
and $q_{\pm}^{ij}$ are the correlation functions  of the measures
$\mu_{\pm}$. The measure $\mu_0$
satisfies
{\bf S0-S2}.
Further, let us assume, in addition to (\ref{S04}), that
\be\la{S5}
q_\pm^{ii}(z)=0,\,\,\,|z|\geq r_0.
\ee
Then the mixing condition {\bf S3}
follows with 
 $\varphi(r)=0$, $r\geq r_0$.
For instance,   (\re{S04}) and (\re{S5}) hold if we set 
$q_\pm^{ii}(z)=
f(z_1)f(z_2)\cdot\dots\cdot f(z_d)$,
 where $f(z)=
N_0-|z|$ for
$|z|\le N_0$ and $f(z)=0$ for
$|z|> N_0$ with
$N_0:=[r_0/\sqrt d]$
(the integer part).
Then by the direct calculation we obtain
$
\hat f(\theta)=(1-\cos N_0\theta)/(1-\cos\theta),$
$\theta\in T^1,$
and (\ref{S04}) holds.

\subsubsection{Non-Gaussian measures}
Let us choose some odd bounded
nonconstant functions
 $f^0,\,f^1\in C(\R)$.
Define $\mu^*_0$ as the distribution of the random function
$
(f^0(Y^0(x)),  f^1(Y^1(x))),
$
where $(Y^0,Y^1)$ is a random function
with  a Gaussian distribution $\mu_0$
 from the previous
example.
Then {\bf S0}-{\bf S3} hold for $\mu^*_0$
with
corresponding mixing coefficient
$\varphi^*(r)= 0$ for $r\geq r_0$.
Measure $\mu^*_0$
is not Gaussian   if
the functions $f^0$, $f^1$ are bounded and nonconstant.

\setcounter{equation}{0}
\section{Application to Second Law}

We apply Theorem~A
to the case when $\mu_{\pm}$ are the  Gibbs measures 
corresponding to
distinct positive
temperatures $T_-\not= T_+$.
We deduce that for the limit
mean energy current
$j_{\infty}=(j_{\infty}^1 ,...,j_{\infty }^d)$
we have
$j_{\infty }^d =-C(T_+-T_-)$ with $C>0$.
Moreover, under the additional condition
on $V$ we obtain $j_{\infty }^k=0$, 
$k=1,\dots,d-1.$
This means that the mean energy current
is directed from high to
low temperature  in accordance with  Second Law.

\subsection{Energy current}
\subsubsection{Energy current for finite energy solutions}
We derive formally
the expression for the
energy current of
the finite energy solutions $u(x,t)$ (see (\ref{H})).
For the half-space $\Omega_k:=\{x\in\Z^d:\,x_k\ge 0\}$
we define the energy in the region
$\Omega_k$ (cf (\ref{H})) as
$$
{\cal E}_k(t):=
\frac12\sum\limits_{x\in\Omega_k}
\Big\{|\dot u(x,t)|^2+\sum\limits_{y\in\Z^d}
\Big(u(x,t),V(x-y)u(y,t)\Big)\Big\}.
$$
By formal
 calculation, using Eqn (\ref{1.1'})
 we obtain
\beqn\la{E_k(t)}
\dot {\cal E}_k(t)=
\frac12\left(
\sum\limits_{x\in\Omega_k^c,\,y\in\Omega_k}
\Big(\dot u(x,t),V(x\!-\!y)u(y,t)\Big)-
\sum\limits_{x\in\Omega_k,\,y\in\Omega_k^c}
\Big(\dot u(x,t),V(x\!-\!y)u(y,t)\Big)
\right).
\eeqn
Here $\Om_k^c:=\Z^d\setminus\Om_k=\{x\in\Z^d: x_k<0\}$.
Introduce new variables:
$x=x'+me_k$, $y=y'+pe_k$,
where $x',y'\in\Z^d$ with $x'_k=y'_k=0$,
$e_k=(\de_{k1},\dots,\de_{kd})$,
$k=1,\dots,d$.
Then we rewrite (\ref{E_k(t)})
in the form
\beqn
\dot {\cal E}_k(t)&=&
\frac12
\sum\limits_{x',y'}
\Big\{
\sum\limits_{m\le-1,\,p\ge 0}
\Big(\dot u(x'\!+\!me_k,t),
V(x'\!+\!me_k \!-\!y'\!-\!pe_k)u(y'\!+\!pe_k,t)
\Big)\nonumber\\
&&-
\sum\limits_{m\ge0,\,p\le -1}
\Big(\dot u(x'\!+\!me_k,t),
V(x'\!+\!me_k\!-\!y'\!-\!pe_k)u(y'\!+\!pe_k,t)
\Big)\Big\}
=\sum\limits_{x'}j^k(x',t).
\nonumber
\eeqn
Here $j^k(x',t)$
stands for the  energy current density
in the direction $e_k$:
by definition,
\beqn
 j^k(x',t):&=&\fr12
\sum\limits_{y'}
\Big\{
\sum\limits_{m\le-1,\,p\ge 0}
\Big(\dot u(x'+me_k,t),V(x'+me_k-y'-pe_k)u(y'+pe_k,t)\Big)
\nonumber\\
&&-
\sum\limits_{m\ge0,\,p\le -1}
\Big(\dot u(x'+me_k,t),
V(x'+me_k-y'-pe_k)u(y'+pe_k,t)\Big)
\Big\},
\nonumber
\eeqn
where $x',y'\in\Z^d$ with $x'_k=y'_k=0$.
\subsubsection{Limit mean energy current}
Now let $u(x,t)$ be the random solution to (\ref{1.1'}) with
the initial measure $\mu_0$ satisfying {\bf S0}--{\bf S3}.
Then  the
bounds  {\bf E1} and
(\ref{sup}) (see below)
imply for the mathematical expectation:
\beqn
E j^k(x',t)&=&
\fr12 \sum\limits_{y'}
\Big(\sum\limits_{m\le-1,\,p\ge 0}
(Q^{10}_t)_{\al\beta} (x+me_k,y'+pe_k) V_{\al\beta}(x'-y'+(m\!-\!p)e_k)
\nonumber\\
&&-
\sum\limits_{m\ge0,\,p\le -1}
(Q^{10}_t)_{\al\beta} (x'+me_k,y'+pe_k) V_{\al\beta}(x'-y'+(m\!-\!p)e_k)
\Big).
\nonumber
\eeqn
Here we omit the summation  on repeating indices $\al,\beta\in\ov n$.
Therefore, from
the  convergence   (\ref{corf})
it follows that
in the limit $t\to\infty$ we get
\beqn
E j^k(x',t)\to   j^k_\infty
&=&\fr12
\sum\limits_{y'}
\Big(
\sum\limits_{m\le-1,\,p\ge 0}
(q^{10}_\infty)_{\al\beta} (x'-y'+(m\!-\!p)e_k) V_{\al\beta}(x'-y'+(m\!-\!p)e_k)
\nonumber\\
&&-
\sum\limits_{m\ge0,\,p\le -1}
(q^{10}_\infty)_{\al\beta} (x'-y'+(m\!-\!p)e_k) V_{\al\beta}(x'-y'+(m\!-\!p)e_k)
\Big).
\nonumber
\eeqn
Denote by
$x'-y'=:z',$ $m-p=:s$
and changing the order of the summation in the
series we get
\beqn
  j^k_\infty
&=&-\fr12
\sum\limits_{z'}\sum\limits_{s\in \Z^1}
(q^{10}_\infty)_{\al\beta} (z'+se_k) V_{\al\beta}(z'+se_k)s=
-\fr12
\sum\limits_{z\in \Z^d}
(q^{10}_\infty)_{\al\beta} (z) z_k V_{\al\beta}(z)
\nonumber\\
&=&
-i\fr{(2\pi)^{-d}}{2}
\int\limits_{T^d}
(\hat q^{10}_\infty)_{\al\beta} (\theta) \pa_k
\ov{\hat V_{\al\beta}(\theta)}\,d\theta,
\quad k=1,\dots,d.\la{jinfty}
\eeqn

\subsection{Gibbs measures}
\subsubsection{Definition of the Gibbs measures}

Formally Gibbs measures $g_\pm$ are 
$$g_{\pm}(du_0, dv_0)= \frac{1}{Z_\pm}
\ds e^{-\ds
 \frac{\beta_{\pm}}{2}\ds
\sum_{x}
(|v_0(x)|^2
+\langle {\cal V}u_0,u_0\rangle)}\prod_{x}
du_0(x)dv_0(x),
$$
where $\beta_{\pm}= T^{-1}_{\pm}$,
$T_\pm\ge0$ are the corresponding absolute temperatures.
We introduce the Gibbs measures $g_{\pm}$ as the
Gaussian measures with
the correlation matrices 
defined by their Fourier transform as
\be\la{4}
\hat q_{\pm}^{00}(\theta)= T_{\pm}\hat V^{-1}(\theta),~~~
\hat q_{\pm}^{11}(\theta)= T_{\pm} 
\Big(\delta_{kl}\Big)_{k,l\in\ov n},~~~
\hat q_{\pm}^{01}(\theta)= \hat q_{\pm}^{10}(\theta)= 0.
\ee
Let $H_\al(\Z^d)$ be the Banach space of the vector-valued
functions $u(x)\in \R^n$ with the finite norm
$$
\Vert u\Vert^2_{\al}\equiv
\sum\limits_{x\in\Z^d}(1+|x|^2)^\al |u(x)|^2<\infty.
$$
Let us fix arbitrary $\alpha<-d/2$.
Introduce   the Gaussian Borel probability  measures
$g_{\pm}^0(du)$, $g_{\pm}^1(dv)$  in spaces
$ H_{\al}(\Z^d)$
 with characteristic
functionals ($\beta_\pm=1/T_\pm$)
$$
\ba{c}
\left.
\ba{rcl}
\hat g_{\pm}^0(\psi)&=&
\ds \int \ds \exp\{i\langle
 u,\psi\rangle\}\,g_{\pm}^0(du)=
\ds \exp\{-\frac{\langle {\cal V}^{-1}\psi,\psi\rangle}{2\beta_\pm}\}\\
~&&\\
\hat g_{\pm}^1(\psi)&=&
\ds \int \ds \exp\{i\langle v,\psi\rangle\}\,g_{\pm}^1(dv)=
\ds \exp\{-\frac{\langle\psi,\psi\rangle}{2\beta_\pm}\}
\ea\right|\,\,\,\,\psi\in D\equiv C_0(\Z^d)\otimes\R^n.
\ea
$$
By the Minlos theorem, \cite{CFS},
the  Borel probability measures $g^0_{\pm}$, $g^1_{\pm}$
exist in the spaces $ H_{\al}(\Z^d)$
 because {\it formally}  we have
\beqn
\int \Vert u\Vert^2_{\alpha}\, g_{\pm}^0(du)
=
\sum\limits_{x\in\Z^d}(1\!+\!|x|^2)^\al
\sum\limits_{i=1}^n \int
u_i(x)u_i(x)\, g_{\pm}^0(du)=
 \sum\limits_{x\in\Z^d}(1\!+\!|x|^2)^\al\,
\tr q^{00}_{\pm}(0)<\infty, \nonumber
\eeqn
since $\alpha<-d/2$ and
\beqn \nonumber
\tr q^{00}_{\pm}(0)=
(2\pi)^{-d}\int\limits_{T^d}\tr
\hat q^{00}_{\pm}(\theta)\,d\theta=
T_{\pm}(2\pi)^{-d}\int\limits_{T^d}\tr \hat V^{-1}(\theta)\,d\theta<\infty.
\eeqn
The last bound is obvious if ${\cal C}_0=\emptyset$
and it follows from condition {\bf E6} if
${\cal C}_0\not=\emptyset$.
Similarly,
\beqn
~~\int\Vert v\Vert^2_{\alpha}\, g_{\pm}^1(dv)
&=&T_{\pm}\, n \sum\limits_{x\in\Z^d}(1+|x|^2)^\al
<\infty,~~\alpha<-d/2.
\nonumber
\eeqn
Finally, we define the Gibbs measures
$g_{\pm}(dY)$ as the Borel probability measures
$g^0_{\pm}(du)\times g^1_{\pm}(dv)$
 in
$\{Y\in{\cal H}_{\al}:Y=(u,v)\}$.
 Let $g_0(dY)$  be a ``two-temperature''
 Borel probability measure in
${\cal H}_{\al}$
that is  constructed  in
  section 2.5.2 with
$\mu_{\pm}(dY)=g_{\pm}(dY)$
and 
 $Y_0$ be a random function with distribution $g_0$. Denote by $g_t$ the
distribution
of $U(t)Y_0$, $t\in\R$.
Now we assume, in addition, that
 ${\cal C}_0=\emptyset$,
i.e. (cf condition {\bf E6})
\be\la{E6'}
\det \hat V(\theta)\not=0,\quad\forall\theta\in T^d.
\ee
Note that in the case of canonical Gibbs
measures
condition {\bf E5'}
 is fulfilled (see Remark~\ref{remi-iii} {\it iii)}).
Indeed, by (\ref{4}) we have
\beqn
\la{Gc00}
p_{kl}^{00}(\theta)&\equiv&
\Big(B^*(\theta) 
\hat q_\pm^{00}(\theta)B(\theta)\Big)_{kl}=
T_\pm\Big(B^*(\theta) \hat V^{-
1}(\theta)B(\theta)\Big)_{kl}=T_\pm\om_k^{-2}(\theta)\de_{kl},
\\
\la{Gc11}
p_{kl}^{11}(\theta)&\equiv&
\Big(B^*(\theta) \hat q_\pm^{11}(\theta)B(\theta)\Big)_{kl}=T_\pm\de_{kl}.
\eeqn
Hence, $p_{kl}^{ij}(\theta)\equiv
\Big(B^*(\theta) \hat q_\pm^{ij}(\theta)B(\theta)\Big)_{kl}=0$ for
$k\not=l$, $\forall i,j$.
\begin{theorem}\la{t1.2'}
Let conditions
 {\bf E1-E4}, (\ref{E6'}) hold and $\al<-{d}/{2}$. Then
there exists
a Gaussian  Borel probability measure  $g_\infty$ on
 ${\cal H}_{\al}$ such that
\be\la{1.8g}
 g_t \,
{\buildrel {\hspace{2mm}{\cal H}_{\al }}\over
{- \hspace{-2mm} \rightharpoondown }}\,
   g_\infty,\,\,\,\,t\to \infty.
\ee
\end{theorem}
{\bf Proof}\,
Let us denote by
 $Q_t(x,y)$
the covariance matrix of measure $g_t$, $t\in\R$.
Note that owing to (\ref{q1}), the matrix
$Q_0(x,y)$
is a ``linear combination''
of $q_\pm(x-y)$.
Hence, $Q_0(x,y)$ satisfies conditions {\bf S0-S2}.
Therefore, by (\ref{4}) we have
$$
|Q_0(x,y)|\le C_1+\sum\limits_\pm
C_\pm|q_\pm^{00}(x-y)|,\,\,\,x,y\in\Z^d.
$$
Condition (\ref{E6'}) implies
\be\la{ft}
|q_\pm^{00}(z)|=T_\pm|F^{-1}_{\theta\to z}
[\hat V^{-1}(\theta)]|\sim (1+|z|)^{-N},\,\,\,\,\forall N\in\N.
\ee
Hence, Lemma \ref{lcom}
 and  Proposition  \re{p4.1}
(with condition {\bf E5'} instead of {\bf E5},
see Remark \ref{remi-iii} {\it iii)})  are
 applicable to
 the correlation matrix
 $Q_t(x,y)$,
  since the proof uses only
the bounds of covariance
(\ref{pr1}), (\ref{pr2}).
These bounds are now provided by
the decay (\ref{ft}) instead of mixing condition
{\bf S3}.
Hence,
$Q_t(x,y)\to Q_\infty(x,y)$,
as $t\to\infty$,
and the family of measures $\{g_t,t\in\R\}$
is weakly compact in ${\cal H}_\al$, $\al<-d/2$.
Hence, 
the convergence (\ref{1.8g})  holds
because  $g_t$ are Gaussian measures.
\hfill$\Box$
\subsubsection{Limit covariance and
energy current for the Gibbs measures}
Now  we rewrite  the
limit covariance 
$\hat q_\infty(\theta)$
and  the limit mean energy current
$j_\infty$ defined by (\ref{jinfty}) 
in the case of
the initial measure  $\mu_0=g_0$ with
$\mu_\pm=g_\pm$ defined above.
At first, by
 (\ref{Pi+})--(\ref{C(theta)}) 
and  (\ref{4})
 we have
\beqn
 M_0^+(\theta)=
\ov T
\left(
\ba{cc}
\hat V^{-1}(\theta)&0\\
0&1
\ea
\right),\quad
 M_0^-(\theta)=
\De T\,
\left(
\ba{cc}
0& \Om^{-1}(\theta)\\
-\Om^{-1}(\theta)&0
\ea
\right),
\la{Q10}
\eeqn
where 
$\ov T:=\ds\frac{T_+\!+\!T_-}{2}$,
$\De T:=\ds\frac{T_+\!-\!T_-}{2}$.
Then, due to (\ref{diag}) and
  (\ref{Piinfty}) we obtain
\beqn
{M_{\infty}^{00}(\theta)}_{kl}
&=&
\ov T\, \om_k^{-2}(\theta)\de_{kl},\,\,\,\,
{M_{\infty}^{11}(\theta)}_{kl}=
\ov T\, \de_{kl},
\nonumber\\
{M_{\infty}^{10}(\theta)}_{kl}&=&-
{M_{\infty}^{01}(\theta)}_{kl}=
-i~\De T\,
\sgn \Big(\frac{\pa\om_k}{\pa\theta_d}(\theta)\Big)
\om_k^{-1}(\theta)\delta_{kl},
\,\,\,k,l\in \ov n.
\nonumber
\eeqn
Therefore,
from (\ref{qinfty}) we get
\beqn\la{34}
\hat q_{\infty}^{00}(\theta)&=&
\ov T\,\hat V^{-1}(\theta),\,\,\,\,
\hat q_{\infty}^{11}(\theta)=\ov T,
\nonumber\\
\hat q_{\infty}^{10}(\theta)
&=&-\hat q_{\infty}^{01}(\theta)
=-i\,\De T\,
B(\theta)\Big[
\sgn \Big(\frac{\pa\om_k}{\pa\theta_d}(\theta)\Big)
\om_k^{-1}(\theta)\delta_{kl}\Big]_{k,l\in\ov n}B^*(\theta).
\eeqn
Substituting $\hat q_{\infty}^{10}(\theta)$
from (\ref{34}) in the RHS of (\ref{jinfty}), we obtain
\beqn
  j^k_\infty &=&
-\fr{\De T}{2(2\pi)^{d}}
\int\limits_{T^d}
B_{\al\de}(\theta)
\sgn \Big(\frac{\pa\om_\de}{\pa\theta_d}(\theta)\Big)
\om_\de^{-1}(\theta)B_{\de\beta}^{*}(\theta)
\pa_k\Big(\ov{B_{\al\gamma}}(\theta)
\om^2_\gamma(\theta)\ov{B_{\gamma\beta}^{*}}(\theta)\Big)
\,d\theta.\,\,\,
\la{jkinfty}
\eeqn
Here as before
we omit the summation  on repeating indices $\al,\beta,\gamma, \de\in\ov n$.
Since  $B(\theta)$ is  the unitary matrix, 
we get
\be\la{mecd}
 j^k_\infty =-
\fr{\De T}{(2\pi)^{d}}
\sum\limits_{\gamma\in\ov n}\,
\int\limits_{T^d}
\sgn \Big(\frac{\pa\om_\gamma}{\pa\theta_d}(\theta)\Big)
\frac{\pa\om_\gamma}{\pa\theta_k}(\theta)
\,d\theta,\quad k=1,\dots,d.
\ee
\begin{remark}\la{rema}
{\it i)}
{\em
From (\ref{mecd})
it follows that
$
j^d_\infty
=-
\ds\fr{\De T}{(2\pi)^{d}}
\sum\limits_{\gamma\in\ov n}\,
\int\limits_{T^d}
\Big|
\frac{\pa\om_\gamma}{\pa\theta_d}(\theta)
\Big|\,d\theta\!<\!0$
if $T_+\!>\!T_-$.\\
{\it ii)}
In  some particular cases we have
$j^k_\infty=0$ for $k=1,\ldots,d-1$:
for example,  a) if  each
$\om_\gamma(\theta)$ is
 even on every variable $\theta_1,\ldots,\theta_{d-1}$,
or b)  if $\ds\sgn \Big(\frac{\pa\om_\gamma}{\pa\theta_d}(\theta)\Big)$
depends only on variable $\theta_d$.
For instance, a) and b) hold
for the simple elastic lattice,
what follows by  (\ref{omega}).\\
{\it iii)} $\hat q^{10}_\infty$ generally is
a discontinuous function  by (\ref{34}).
Therefore, $ q^{10}_\infty(x)$ decays as a negative power
of $|x|$. The exponential decay is impossible in contrast
with the case
of \ci{Na}.
}
\end{remark}

\setcounter{equation}{0}
\section{Bounds for initial covariance}

\begin{definition}
By $l^p\equiv l^p(\Z^d)\otimes \R^n$, $p\ge 1$,
$n\ge 1$,
 we denote the space of sequences
$f(k)=(f_1(k),\dots,f_n(k))$ endowed with norm
$\Vert f\Vert_{l^p}=\Big(\sum\limits_{k\in\Z^d}|f(k)|^p\Big)^{1/p}$.
\end{definition}
The next Proposition reflects
 the mixing property in the
Fourier transforms
$\hat q^{ij}_\pm$
of initial correlation functions
$q^{ij}_\pm$.
Condition
 {\bf S2} implies that  $q^{ij}_\pm(z)$
are   bounded  functions.
Therefore, its  Fourier transform
 generally belongs to the Schwartz space
of tempered distributions.
\begin{pro} \la{l4.1}
Let  conditions {\bf S0-S3} hold.
Then\\
i) For $i,j=0,1,$
the following bounds hold
\beqn
\sum\limits_{y\in\Z^d} |Q^{ij}_0(x,y)|
&\le& C<\infty\,\,\,\mbox{ for all }\,x\in\Z^d,
\la{pr1}
\\
\sum\limits_{x\in\Z^d} |Q^{ij}_0(x,y)|
&\le& C<\infty\,\,\,\mbox{ for all }\,y\in\Z^d.
\la{pr2}
\eeqn
Here the constant $C$
does not depend on $x,y\in \Z^d$.\\
ii) $\hat q^{ij}_\pm\in  C(T^d)$,~ $i,j=0,1$.
\end{pro}
{\bf Proof}
{\it ad i)}
Conditions {\bf S0},
{\bf S2} and {\bf S3} imply by
\ci[Lemma 17.2.3]{IL} (or Lemma \re{il} i) below):
\be\la{4.9'}
|Q^{ij}_0(x,y)|
\le C e_0\,\varphi^{1/2}(|x-y|),~~ x,y\in\Z^d.
\ee
Hence,
 (\ref{1.12}) implies (\ref{pr1}):
\be\la{qp}
\sum\limits_{y\in\Z^d}
|Q^{ij}_0(x,y)| \le C e_0
 \sum\limits_{z\in\Z^d} \varphi^{1/2}(|z|) <\infty.
\ee
{\it ad ii)}
The bound  (\ref{4.9'})  and condition
(\ref{3a}) imply
the following bound:
\be\la{4.9}
|q^{ij}_\pm(z)|\le C e_0\,\varphi^{1/2}(|z|),~~ z\in\Z^d.
\ee
Hence, from (\ref{1.12})
it follows that $q^{ij}_\pm(z)\in l^1$,
what  implies $\hat q^{ij}_\pm\in C(T^d)$.
\hfill$\Box$

\begin{cor}\la{c4.1}
Proposition \ref{l4.1}, i)
implies, by the Shur lemma,
 that for any
$\Phi,\Psi\in l^2$
the following bound holds:
\be
|\langle
Q_0(x,y),\Phi(x)\otimes\Psi(y)\rangle|\le
C\Vert\Phi\Vert_{l^2}
\Vert\Psi\Vert_{l^2}.
\ee
\end{cor}

\setcounter{equation}{0}
\section{Compactness of measures family}

Proposition \re{l2.1}  follows
 from the bound (\re{20.1})
by the
Prokhorov Theorem \cite[Lemma II.3.1]{VF}
using the method of 
\ci[Theorem XII.5.2]{VF},
since the embedding 
${\cal H}_\al\subset {\cal H}_\beta$
is compact if $\al>\beta$.
\begin{lemma}\la{lcom}
Let conditions {\bf S0, S2, S3} hold and
$\al<-d/2$.
Then  the following bounds hold
 \beqn
\sup\limits_{t\ge 0}
E\Vert U(t)Y_0\Vert^2_\al<\infty.      \la{20.1}
\eeqn
\end{lemma}
{\bf Proof}\,
 Definition (\re{d1.1})
 implies
\be
E \Vert  Y(\cdot,t)\Vert^2_\al=
\!\sum\limits_{x\in \Z^d}
(1+|x|^2)^\al
\Big({\rm tr}\,Q_t^{00}(x,x)+
{\rm tr}\,Q_t^{11}(x,x)\Big)<\infty.~~~\la{E0Q}
\ee
Since  $\al<-d/2$,
it remains to prove that
\be \la{sup}
\sup\limits_{t\in\R} \sup\limits_{x,y\in \Z^d}
\Vert Q_t(x,y)\Vert\le C<\infty.
\ee
The representation (\re{solGr})
gives
\beqn
Q^{ij}_t(x,y)&=&
E\Big(Y^i(x,t)\otimes Y^j(y,t)\Big)
=
\sum\limits_{x',y'\in \Z^d}
\sum\limits_{k,l=0,1}
{\cal G}^{ik}_t(x\!-\!x')Q^{kl}_0(x',y'){\cal G}^{jl}_t(y\!-\!y')\nonumber\\
&=&
\langle Q_0(x',y'), \Phi^i_{x}(x',t)\otimes
\Phi^j_{y}(y',t)\rangle,
\eeqn
where
$$
\Phi^i_{x}(x',t):=\Big(
{\cal G}^{i0}_t(x-x'),{\cal G}^{i1}_t(x-x')\Big),\,\,
\,\,\,x'\in\Z^d,\,\,\,\,\,\,i=0,1.
$$
Note that the
Parseval identity, (\ref{hatcalG})
and condition {\bf E6} imply
$$
\Vert\Phi^i_{x}(\cdot,t)\Vert^2_{l^2}= (2\pi)^{-d}
\int\limits_{T^d}
|\hat\Phi^i_{x}(\theta,t)|^2\,d\theta
=(2\pi)^{-d}\int\limits_{T^d}
\Big(
|\hat{\cal G}^{i0}_t(\theta)|^2
+|\hat{\cal G}^{i1}_t(\theta)|^2\Big)
\,d\theta
\le C_0<\infty.
$$
Then Corollary \ref{c4.1} gives
\be
|Q^{ij}_t(x,y)|=
|\langle Q_0(x',y'), \Phi^i_{x}(x',t)\otimes
\Phi^j_{y}(y',t)\rangle|
\le C\Vert\Phi^i_{x}(\cdot,t)\Vert_{l^2}\,
\Vert\Phi^j_{y}(\cdot,t)\Vert_{l^2}
\le C_1<\infty,
\la{5.4}
\ee
where the constant $ C_1$  does  not depend on
$x,y\in\Z^d$, $t\in\R$.
\hfill$\Box$

\setcounter{equation}{0}
\section{``Cutting out'' of critical spectrum}
We reduce the proof of the convergences (\ref{corf}) and
 (\ref{2.6i}) by a suitable spectral analysis.
\subsection{Equicontinuity of covariance}
Obviously, (\ref{corf}) is equivalent to the next proposition.
\begin{pro}\la{p4.1}
Let conditions
{\bf E1}-{\bf E6} and {\bf S0}-{\bf S3} hold.
Then $\forall \Psi\in {\cal D}$
\be\la{4.4}
{\cal Q}_t(\Psi,\Psi)
\to {\cal Q}_\infty(\Psi,\Psi),\,\,\,\,t\to \infty.
\ee
\end{pro}

Let us show that we can restrict ourselves
$\Psi\in {\cal D}^0$, where
${\cal D}^0$ is a subset of functions $\Psi\in{\cal D}$
with
vanishing spectrum in a neighborhood of a critical set ${\cal C}\subset T^d$.
For $k=1,...,n$ define the sets
$$
Z_k:=\{\theta\in T^d\setminus {\cal C}_*:
~\nabla_{\theta_d}\om_k(\theta)=0\}.
$$
\begin{definition}\la{dC}
i) The critical set ${\cal C}:={\cal C}_0\cup{\cal C}_*
\cup\Big(\cup_1^n Z_k\Big)\cup\Big(\cup_1^n{\cal C}_k\Big) $
(see {\bf E4}).
\\
ii)
${\cal D}^0:=\{\Psi\in{\cal D}:
\hat\Psi(\theta)=0\quad \mbox{\rm in a neighborhood of}\quad{\cal C}
\}$.
\end{definition}
The next lemma  plays the central role
in our arguments although its proof
is similar to the proofs
of  Lemmas \ref{lc*} and \ref{lc}
since ${\cal C}\ne T^d$.
\begin{lemma}\la{mesF} Let conditions {\bf E1-E4} hold. Then
{\rm mes}\,${\cal C}=0$.
\end{lemma}

Next, we introduce a norm $\Vert\cdot\Vert_V$ in the space  ${\cal D}$
such that i) ${\cal D}^0$ is dense in  ${\cal D}$ in this norm, while
ii) the quadratic forms ${\cal Q}_t(\Psi,\Psi)$, $t\in\R$, are
{\it equicontinuous} in this norm. Then it suffices to prove
 (\ref{4.4}) for $\Psi\in {\cal D}^0$ only.
\begin{definition}\la{dDV}
 ${\cal D}_V$ is the space  ${\cal D}$ endowed with the norm
\be
\Vert\Psi\Vert_V^2:=
\int\limits_{T^d}
(1+\Vert V^{-1}(\theta)\Vert)|\hat\Psi(\theta)|^2
\,d\theta,\quad \Psi\in{\cal D},\la{nDV}
\ee
which is  finite by condition {\bf E6}.
\end{definition}
The set ${\cal D}^0$ is dense in  ${\cal D}_V$
by Lemma \re{mesF} and condition {\bf E6}.
\begin{lemma}\la{lcc}
The quadratic forms  ${\cal Q}_t(\Psi,\Psi)$, $t\in\R$, are
equicontinuous
in ${\cal D}_V$.
\end{lemma}
{\bf Proof}\, It suffices to prove the uniform bounds
\be
\sup\limits_{t\in\R} |{\cal Q}_t(\Psi,\Psi)|\le C
\Vert\Psi\Vert_V^2,\quad \Psi\in{\cal D}.
\la{cub}
\ee
 Definition (\re{qd}) implies that
$ {\cal Q}_t(\Psi,\Psi):=E|\langle Y(x,t),\Psi(x)\rangle|^2$.
Note that 
\beqn\la{YP}
\langle Y(x,t),\Psi(x)\rangle
=\langle Y_0(x),\Phi(x,t)\rangle,
\eeqn
where $\Phi(\cdot,t):=
F^{-1}[\hat{\cal G}^*_t(\theta)\hat \Psi(\theta)]$.
Therefore,
$ {\cal Q}_t(\Psi,\Psi)
={\cal Q}_0(\Phi(\cdot,t),\Phi(\cdot,t))
$, so
\be
\sup\limits_{t\in\R} |{\cal Q}_t(\Psi,\Psi)|\le C
\sup\limits_{t\in\R} \Vert\Phi(\cdot,t)\Vert_{l^2}^2
\la{cubc}
\ee
by Corollary \ref{c4.1}.
Finally,
by the Parseval identity and (\ref{hatcalG}), we get
\be\la{esPhi}
\,\,\,\,\,\,\,\,\,\,\Vert\Phi(\cdot,t)\Vert_{l^2}^2=
(2\pi)^{-d}\int_{T^d}
\Vert \hat {\cal G}_t^*(\theta)\Vert^2 |\hat\Psi(\theta)|^2 d\theta \le C
\Vert\Psi\Vert_V^2.
\,\,\,\,\,\,\,\,\,\,\,\,\,\,\,\,\,\,\,\,{\Box}
\ee
\subsection{Equicontinuity
 of characteristic functionals}
The convergence (\ref{2.6i}) also it suffices
to prove for $\Psi\in{\cal D}^0$ only.
This follows from the next lemma.
\begin{lemma}\la{lck}
The characteristic functionals
$\hat\mu_t(\Psi)$, $t\in\R$, are
equicontinuous
in ${\cal D}_V$.
\end{lemma}
{\bf Proof}\, This lemma follows immediately from 
 Lemma \re{lcc}
by the Cauchy-Schwartz inequality:
$$
\ba{rcl}
|\hat\mu_t(\Psi_1)-\hat\mu_t(\Psi_2)|&=&
|\ds\int \Big( e^{i\langle Y,\Psi_1 \rangle}-
e^{i\langle Y,\Psi_2 \rangle}\Big)\mu_t(dY)|
\le
\ds\int |e^{i\langle Y,\Psi_1-\Psi_2 \rangle}-1|\mu_t(dY)\\
~\\
&\le&
\ds\int |\langle Y,\Psi_1-\Psi_2 \rangle|\mu_t(dY)
\le \sqrt {\ds\int |\langle Y,\Psi_1-\Psi_2 \rangle|^2
\mu_t(dY)}\\
~\\
&=&
\sqrt {{\cal Q}_t(\Psi_1-\Psi_2, \Psi_1-\Psi_2)}
\le C\Vert\Psi_1-\Psi_2 \Vert_V.
~~~~~~~~~~\Box
\ea
$$

\setcounter{equation}{0}
\section{Convergence of covariance for non-critical spectrum}
We prove Proposition
\re{p4.1} for  $\Psi\in {\cal D}^0$.
First we split the initial covariance into the following matrices
\beqn
Q^+(x,y)&:=&{\bf q^+}(x-y),\la{d1'}\\
Q^-(x,y)&:=&{\bf q^-}(x-y)\sgn y_d,\la{d1''}\\
Q^r(x,y)&:=&Q_0(x,y)-Q^+(x,y)-Q^-(x,y).\la{d1'''}
\eeqn
where
${\bf q^+}=\ds\frac{1}{2} (q_++q_-)$,
${\bf q^-}=\ds\frac{1}{2} (q_+-q_-)$.
Since the solution $Y(t)$ to Cauchy problem
 (\ref{1.1'}) admits the representation
(\ref{solGr}), we have
$$
Q_{t}(x,y)=
\sum\limits_{x',y'\in\Z^d}
\Big({\cal G}_t(x\!-\!x') Q_0(x',y')
{\cal G}_t^T(y\!-\!y')\Big).
$$
Next introduce the  matrices
\be\la{Qta}
Q^a_{t}(x,y)=
\sum\limits_{x',y'\in\Z^d}
\Big({\cal G}_t(x\!-\!x') Q^a(x',y')
{\cal G}_t^T(y\!-\!y')\Big),\,\,\,\,
x,y\in \Z^d,\,\,\,\,t>0,
\ee
for each $a=\{+,-,r\}$,
and split
$Q_t(x,y)$ into three terms:
$Q_t(x,y)= Q^+_{t}(x,y)+Q^-_{t}(x,y)+Q^r_{t}(x,y)$.
Below in Lemmas \ref{Qt1}, \ref{Qt3}, \ref{Qt4}
 we will
prove the convergence of type (\ref{p4.1}) 
to a limit for each
term $Q^a_{t}(x,y)$.

\subsection{Convergence of $Q^+_t(x,y)$}
\begin{lemma}\la{Qt1}
$\lim\limits_{t\to\infty}\langle Q_t^+(x,y),\Psi(x)\otimes\Psi(y)\rangle
= \langle q^+_\infty(x-y),\Psi(x)\otimes\Psi(y)\rangle$
for any 
$ \Psi\in {\cal D}^0$,
where the matrix
$q^+_{\infty}$ is defined by (\ref{qinfty+}).
\end{lemma}
{\bf Proof}
At first,
let us apply the Fourier transform to the matrix
$Q_{t}^+(x,y)$
defined by (\ref{Qta}). Then we have
$\hat Q^{+}_{t}(\theta,\theta'):=
F\!\!\!_{\scriptsize {\ba{ll}
x\to\theta\\ y\to -\theta'
\ea}}\!\! Q^{+}_{t}(x,y)=
\hat {\cal G}_t(\theta)\hat Q^{+}
(\theta,\theta')\hat {\cal G}_t^T(-\theta'),
$
where
$\hat Q^{+}
(\theta,\theta'):=F\!\!\!_{\scriptsize {\ba{ll}
x\to\theta\\ y\to -\theta'
\ea}}\!\! Q^+(x,y)$.
From  (\ref{d1'})
it follows that
$
\hat Q^+(\theta,\theta')=
\delta(\theta-\theta')~(2\pi)^{d}~
\hat{\bf q}^+(\theta).
$
Hence, 
\be\la{FtQt+}
\hat Q^+_{t}(\theta,\theta')=(2\pi)^d\delta(\theta-\theta')
\hat {\cal G}_t(\theta)
\hat{\q}^+(\theta) \hat{\cal G}_t^*(\theta).
\ee
Here we use that 
$\hat {\cal G}_t^T(-\theta)=
\hat {\cal G}_t^*(\theta)$ 
by condition {\bf E2}. Therefore,
\beqn\la{Qt,1}
\langle Q_t^+(x,y),\Psi(x)\otimes\Psi(y)\rangle
&=&(2\pi)^{-2d}
\langle \hat Q_t^+(\theta,\theta'),\hat \Psi(\theta)
\otimes\ov{\hat\Psi}(\theta')\rangle
\nonumber\\
&=&
(2\pi)^{-d}\langle
\hat {\cal G}_t(\theta)
\hat \q^+(\theta)\hat {\cal G}_t^*(\theta),
\hat \Psi(\theta)\otimes\ov{\hat \Psi}(\theta)\rangle.
\eeqn
Further,
we choose certain smooth branches of the functions $B(\theta)$ and $\om_k(\theta)$ to apply the stationary phase arguments which
require a smoothness in $\theta$.
We  choose
 a finite partition of unity 
\be\la{part} 
\sum_{m=1}^M g_m(\theta)=1,\,\,\,\,\theta\in \supp
\hat\Psi, 
\ee 
where $g_m$ are nonnegative functions from 
$C_0^\infty(T^d)$ and    
vanish  in a neighborhood of the set
${\cal C}$
  defined 
in Definition \ref{dC}, i).
Further, using (\ref{part}) we rewrite
the  RHS of (\ref{Qt,1}).
Applying formulas (\ref{Gtdec}),
(\ref{G*tdec}) for $\hat {\cal G}_t(\theta)$,
 $\hat {\cal G}^*_t(\theta)$,
 we obtain (see Appendix)
\be\la{qtpar}
\langle Q_t^+(x,y),\Psi(x)\otimes\Psi(y)\rangle
=(2\pi)^{-d}\sum_{m}
\int\limits_{T^d}  g_m(\theta) 
\Big(B(\theta)R_t(\theta) B^*(\theta),
\hat \Psi(\theta)\otimes\ov{\hat \Psi}(\theta)\Big)\,d\theta,
\ee
where by $R_t(\theta)$ 
we denote the $2n\times 2n$ matrix
with the entries (cf (\ref{ap.2})): 
\beqn\la{7.9}
R_t(\theta)_{kl}&=&\frac{1}{2}\sum\limits_{\pm}\Big\{
\cos\big(
\om_k(\theta)\!\pm\!\om_l(\theta)\big)t
~\Big[
B^*(\theta)
\Big(
\hat \q^+(\theta)\mp\hat C(\theta)\hat \q^+(\theta)\hat
C^*(\theta)\Big)B(\theta)\Big]_{kl}
\nonumber\\
&&+
\sin\big(\om_k(\theta)\!\pm\!\om_l(\theta)\big)t
~\Big[
B^*(\theta)
\Big(
\hat C(\theta)\hat \q^+(\theta)\pm\hat \q^+(\theta)\hat
C^*(\theta)
\Big)B(\theta)\Big]_{kl}\Big\}.
\eeqn 
By Lemma \re{lc*} and the compactness arguments, 
we choose the eigenvalues $\om_k(\theta)$ 
and the matrix $B(\theta)$  
as  
real-analytic functions inside 
the $\supp g_m$ for every $m$: we do not mark the 
functions by the index 
$m$ to not overburden the notations.
Now we analyze the Fourier 
integrals with   $g_m$.

At first, note that the
identitites  
$\om_k(\theta)+\om_l(\theta)\equiv\const_+$ 
or 
$\om_k(\theta)-\om_l(\theta)\equiv\const_-$
 with the $\const_\pm\ne 0$
are impossible by   condition {\bf E5}.
Furthermore, 
the oscillatory integrals 
with $\om_k(\theta)\pm \om_l(\theta)\not\equiv \const$
vanish as $t\to\infty$. Hence,
 only the integrals with  $\om_k(\theta)-\om_l(\theta)\equiv 0$
contribute to the limit,
since  $\om_k(\theta)+\om_l(\theta)\equiv 0$ would imply  
$\om_k(\theta)\equiv\om_l(\theta)\equiv 0$ which 
is impossible by  {\bf E4}.
We enumerate the eigenvalues
$\om_k(\theta)$  as in (\ref{enum}).
Then if $k,l\in(r_{\sigma-1},r_{\sigma}]$, we have
$\cos(\om_k-\om_l)t=1$ for $\sigma=1,\dots,s+1$.
By  formula (\ref{ap.3})
with $\hat q(\theta):=\hat \q^+(\theta)$
and (\ref{qtpar}), (\ref{7.9}), we get
\beqn\la{3.10}
&&
\langle Q_t^+(x,y),\Psi(x)\otimes\Psi(y)\rangle
\nonumber\\
&=&
(2\pi)^{-d}
\sum\limits_m\int\limits_{T^d}
g_m(\theta)\Bigl(B(\theta)
\Big[\chi_{kl}
\Big(B^*(\theta)M_0^{+}(\theta)
B(\theta)\Big)_{kl}\Big]_{k,l\in\bar n}
B^*(\theta)+\dots,
\hat \Psi(\theta)\otimes\ov{\hat \Psi}(\theta)\Big)\,d\theta
\nonumber\\
&=&
(2\pi)^{-d}\int\limits_{T^d}
\Bigl(\hat q^+_{\infty}(\theta),
\hat \Psi(\theta)\otimes\ov{\hat \Psi}(\theta)\Big)\,d\theta +\dots,\,\,\,\,\,\,\,\,\,\,\,\,
\eeqn
where 
$M_0^{+}(\theta)$ is defined in (\ref{Pi+}),
 $"\dots"$ stands for the oscillatory
 integrals
which contain 
 $\cos(\om_k(\theta)\pm\om_l(\theta))t$ 
and $\sin(\om_k(\theta)\pm\om_l(\theta))t$ 
with $\om_k(\theta)\pm\om_l(\theta)\not\equiv$const. 
The oscillatory integrals 
converge to zero  by the Lebesgue-Riemann Theorem  
since all the integrands in `$...$'  are summable, and  
$\na(\om_k(\theta)\pm\om_l(\theta))=0$ only on the set 
of the Lebesgue measure zero. 
The summability follows from 
Proposition \ref{l4.1}, ii)
and  {\bf E6} (if ${\cal C}_0\not=\emptyset$)  
 since 
the matrices $B(\theta)$ are unitary. 
The zero measure follows similarly to (\re{c*}) 
since $\om_k(\theta)\pm\om_l(\theta)\not\equiv$const.  Lemma \ref {Qt1} is proved. 
\hfill$\Box$

\subsection{ Convergence of $Q^-_t(x,y)$}
\begin{lemma}\la{Qt3}
$\lim\limits_{t\to\infty}\langle Q_t^-(x,y),\Psi(x)\otimes\Psi(y)\rangle
= \langle q^-_\infty(x-y),\Psi(x)\otimes\Psi(y)\rangle$
for any $ \Psi\in {\cal D}^0$,
where  the matrix
$ q^-_{\infty}$ is defined in (\ref{qinfty-}).
\end{lemma}
{\bf Proof}
{\it Step 1}\,
At first
 we apply the Fourier transform to
$Q_{t}^-(x,y)$
defined by (\ref{Qta}):
\be\la{hatQt-}
\hat Q^{-}_{t}(\theta,\theta'):=
F\!\!\!_{\scriptsize {\ba{ll}
x\to\theta\\ y\to -\theta'
\ea}}\!\! Q^{-}_{t}(x,y)=
\hat {\cal G}_t(\theta)\hat Q^{-}
(\theta,\theta')\hat {\cal G}_t^T(-\theta'),
\ee
where $\hat Q^{-}
(\theta,\theta'):=F\!\!\!_{\scriptsize {\ba{ll}
x\to\theta\\ y\to -\theta'
\ea}}\!\! Q^{-}(x,y)$.
Similarly to 
(\ref{Qt,1}) and (\ref{qtpar})
using the partition of unity
 (\ref{part})
and also  formulas (\ref{ap.11})
and (\ref{Rkl})
we obtain
\beqn\la{6.20}
&&\langle Q_t^-(x,y),\Psi(x)\otimes\Psi(y)\rangle
=(2\pi)^{-2d}
\langle \hat Q_t^-(\theta,\theta'),\hat \Psi(\theta)
\otimes\ov{\hat\Psi}(\theta')\rangle
\nonumber\\
&=&
(2\pi)^{-2d}\sum\limits_{m,m'}
\langle g_m(\theta)g_{m'}(\theta')
B(\theta)R_t(\theta,\theta') B^*(\theta'),
\hat \Psi(\theta)
\otimes\ov{\hat\Psi}(\theta')\rangle,
\eeqn
where $R_t(\theta,\theta')$
is defined in (\ref{Rkl})
with $\hat Q(\theta,\theta'):=
\hat Q^-(\theta,\theta')$.
Second, we have
$
F_{y\to \theta}(\sgn y)=
i~\ds\PV(\frac {1}{\tg(\theta/2)}),$ $\theta\in T^1,$
where $\PV$ stands for the
Cauchy principal part and $y\in\Z^1$.
Hence, by (\ref{d1''}),
we obtain
\be\la{Q2}
\hat Q^-(\theta,\theta')=
\delta(\bar \theta-
\bar \theta')~(2\pi)^{d-1}~
i~
\PV(\frac {1}{\tg(\theta_d-\theta'_d)/2})
\hat {\bf q}^-(\theta).
\ee
Here
$\bar \theta=(\theta_1,\dots,\theta_{d-1}),\,
 \bar \theta'=(\theta'_1,\dots,\theta'_{d-1}),\,
\theta=(\bar\theta,\theta_{d}),\,
\theta'=(\bar\theta',\theta'_{d})\in T^d$.
Note that the Fourier transform
of $Q_t^-(x,y)$ is more singular than of $Q_t^+(x,y)$
(cf. formulas (\ref{FtQt+}) and (\ref{hatQt-}), (\ref{Q2})).
Therefore it is of key importance that
we can restrict ourselves by the functions
$\Psi\in{\cal D}^0$.
Further, (\ref{Q2})
and (\ref{Rkl}) 
with $\hat Q(\theta,\theta'):=
\hat Q^-(\theta,\theta')$
imply
\beqn\la{6.90}
R_t(\theta,\theta')_{kl}&=& 
\delta(\bar \theta-\bar \theta')
(2\pi)^{d-1}i\,
\PV(\frac{1}{\tg(\theta_d-\theta'_d)/2})
\nonumber\\
&&\cdot\sum\limits_{\pm}
\Big\{
\cos\om^\pm_{kl}t
~\Big(M_1^\pm(\theta,\theta')\Big)_{kl}
+
\sin\om^\pm_{kl}t
~\Big(M_2^\pm(\theta,\theta')\Big)_{kl}\Big\}.
\eeqn
Here 
$\om^\pm_{kl}\equiv
\om^\pm_{kl}(\theta,\theta'):=
\om_k(\theta)\pm\om_l(\theta')$,
$M_1^\pm(\theta,\theta'):=
B^*(\theta)
\ds\frac{1}{2}\Big(\hat {\bf q}^-(\theta)
\mp\hat C(\theta)
\hat {\bf q}^-(\theta)
\hat
C^*(\theta')\Big)B(\theta')$,
$M_2^\pm(\theta,\theta'):=
B^*(\theta)
\ds\frac{1}{2}\Big(\hat C(\theta)
\hat {\bf q}^-(\theta)\pm
\hat {\bf q}^-(\theta)\hat
C^*(\theta')
\Big)B(\theta')$.
Let us analyse the summands in the  RHS of (\ref{6.20}).
Since  
$\cos\big(\om_{kl}^\pm t\big)=
\ds\frac{e^{i\om^\pm_{kl}t}+e^{-i\om^\pm_{kl}t}}{2}$
and
$\sin\big(\om_{kl}^\pm t\big)=
\ds\frac{e^{i\om^\pm_{kl}t}-e^{-i\om^\pm_{kl}t}}{2i}$
it suffices to prove
the convergence 
for  arising integrals $I^{\pm}_{kl}(t)$
resp.  $J^{\pm}_{kl}(t)$
with $e^{i\om^\pm_{kl}t}$
resp. $e^{-i\om^\pm_{kl}t}$ 
 (see  {\it Step 2}  resp. {\it Step 3}).

{\it Step 2}\,
First, we consider the integrals $I^{\pm}_{kl}(t)$.
Let us denote, for simplicity of exposition,
$g_m\equiv g_m(\theta)$,
$g_{m'}\equiv g_{m'}(\theta')$
and   $\hat\Psi_{r}(\theta):=
(\hat\Psi^0_r(\theta), \hat\Psi^1_r(\theta))$.
Also let us denote 
by $p_{kl}(\theta,\theta')$
one of the expressions
 $B_{rk}(\theta)(M_i^\pm(\theta,\theta'))_{kl}
B^*_{ks}(\theta')$
with either $+$ or $-$, and some $i=1,2$, $r,s\in\ov n$.
Then (\ref{6.20}) and (\ref{6.90}) give,
\beqn\la{6.91}
{I^{\pm}_{kl}}(t):&=&
(2\pi)^{-2d}
\langle
g_m g_{m'}\de(\bar\theta\!-\!\bar\theta')(2\pi)^{d-1}i
\PV\frac{1}{\tg(\theta_d\!-\!\theta'_d)/2}
e^{i\om_{kl}^\pm t}
p_{kl}(\theta,\theta'),\hat\Psi_r(\theta)\otimes
\ov{\hat\Psi_s}(\theta')\rangle
\nonumber\\
&=&\!\!\!(2\pi)^{-d-1}i
\int\limits_{T^{d}}
g_m  e^{i\omega_k(\theta)t}\ov{\hat\Psi_{r}}(\theta)
\Big(\PV \int\limits_{T^1} 
g_{m'}\left.
e^{\pm i\omega_l(\theta')t}
\frac{p_{kl}(\theta,\theta')
\hat\Psi_{s}(\theta')}{\tg(\theta_d\!-\!\theta'_d)/2}\right|_{\theta'=(\ov\theta,\theta'_d)}
\!\!\!\!\!\!\!\!d\theta'_d\Big)d\theta.\,\,\,\,\,\,\,\,\,
\eeqn
The integral with $\PV$ in the RHS of (\ref{6.91})
exists since
$\om_l(\theta')$ are analytic inside
the  $\supp g_{m'}(\theta')$.
Changing variables
$\theta'_d\to \theta_d-\theta'_d=\xi$ in the
inner integral in  the RHS of (\ref{6.91})
we obtain
\be\la{6.16}
I^{\pm}_{kl}(t)=
(2\pi)^{\!-\!d\!-\!1}i\int\limits_{T^{d}}
\!g_{m}e^{i\omega_k(\theta)t}
\ov{\hat \Psi_r}(\theta)
\Big(
\PV\int\limits_{T^1}
g_{m'}e^{\pm i\om_l(\theta') t}
\left.\frac{p_{kl}(\theta,\theta')
\hat\Psi_s(\theta')}{\tg(\xi/2)}
\right|_{\theta'=(\ov\theta,\theta_d-\xi)}
\!\!\!d\xi
\Big)d\theta.
\ee
From Definition \ref{dC} 
it follows that $\nabla_{\theta'_d}\om_l(\theta')\not=0$
for $\theta'\in\supp g_{m'}\subset \supp\hat\Psi$.
Next lemma follows from
  \ci[Proposition A.4 i), ii)]{BPT}.
\begin{lemma}\la{auxl}
Let $\na_{\theta_d}\om_l(\theta)
\not=0$ for $\theta\in\supp g_{m'}$
and 
$p(\theta)\in C^1(T^d)$.
Then
\beqn
P_l(\theta,t)&:=&
\PV\int\limits_{T^1}
g_{m'}({\overline \theta},\theta_d-\xi)
\frac{e^{\pm i\om_l({\overline \theta},\theta_d-\xi)t}}
{\tg(\xi/2)}p(\bar \theta,\theta_d-\xi)
\,d\xi\nonumber\\
 &=&2\pi i\,
g_{m'}(\theta)\,
e^{\pm i\om_l(\theta)t}p(\theta)
\sgn(\mp\frac{\pa \om_l}{\pa\theta_d}(\theta))
+o(1),\,\,\,\,t\to+\infty,\la{14.1}
\eeqn
for $\theta\in\supp g_{m'}$, and
\beqn\la{14.2}
\sup\limits_{\theta\in T^d,
t\in\R,l\in\ov n}
 |P_l(\theta,t)|<\infty.
\eeqn
\end{lemma}
 Applying Lemma \ref{auxl} to
the inner 1D integral in
 (\ref{6.16}), we obtain  as $t\to+\infty$,
\beqn\la{6.16'}
I^{\pm}_{kl}(t)=-(2\pi)^{-d}
\int\limits_{T^{d}}\!
g_m(\theta)g_{m'}(\theta) e^{i\omega^\pm_{kl}
(\theta,\theta)t}
p_{kl}(\theta,\theta)
 \sgn\big(\!\mp\! \frac{\partial \om_l}
{\partial \theta_d}(\theta)\big)
\ov{\hat \Psi_r}(\theta)\hat\Psi_s(\theta)\,d\theta
\!+\!o(1),\,\,\,
\eeqn
where $\omega^\pm_{kl}
(\theta,\theta)\equiv
\om_k(\theta)\pm\om_l(\theta)$.
Let us discuss the limits of the integrals
$I_{kl}^\pm(\theta)$ as $t\to+\infty$.
At first, we note that
the identities
$\om^+_{kl}(\theta,\theta)\equiv\const_+$ or 
$\om^-_{kl}(\theta,\theta)\equiv\const_-$
with the $\const_\pm\ne0$
are impossible by condition {\bf E5}.
On the other hand,
the oscillatory integrals with 
$\om^\pm_{kl}(\theta,\theta)\not\equiv\const$
vanish as $t\to\infty$
 owing  to Proposition \ref{l4.1}, ii),
 {\bf E4}, {\bf E5},
  {\bf E6} (if ${\cal C}_0\not=\emptyset$)  
and the Lebesgue-Riemann theorem
(as in Lemma \ref{Qt1}). Hence,
\be\la{6.92}
I^{+}_{kl}(t)
\to 0,\,\,\,t\to+\infty.
\ee
Similarly,
in the case $\om^-_{kl}(\theta,\theta)
\not\equiv0$,
we have
$I^-_{kl}(t)\to 0$, as $t\to\infty$.
Therefore, only integrals with
$\om^-_{kl}(\theta,\theta)\equiv 0$,
i.e.   $k,l\in(r_{\sigma-1},r_\sigma]$,
$\sigma=1,\dots, s+1$
(see (\ref{enum})), contribute to a limit.
Finally, by (\ref{chi}), we get 
\beqn\la{6.93}
I^{-}_{kl}(t)=-(2\pi)^{-d}
\int\limits_{T^{d}}
g_m g_{m'} \chi_{kl}\,
p_{kl}(\theta,\theta)
 \sgn\Big( \frac{\partial \om_k}{\partial \theta_d}(\theta)\Big)
\ov{\hat \Psi_r}(\theta)\hat\Psi_s(\theta)
\,d\theta
+o(1),\, t\to+\infty.
\eeqn
{\it Step 3}\,
Now consider the integrals $J^\pm_{kl}(t)$ 
of type (\ref{6.91})
 with $e^{-i\om^\pm_{kl}t}$ instead
of $e^{i\om^\pm_{kl}t}$.
Similarly to  
 (\ref{6.91})-(\ref{6.92}), we get
\beqn\la{6.13}
J^{+}_{kl}(t):&=&
(2\pi)^{-2d}
\langle
g_m g_{m'}\de(\bar\theta\!-\!\bar\theta')(2\pi)^{d-1}i
\PV\frac{1}{\tg(\theta_d\!-\!\theta'_d)/2}
e^{-i\om_{kl}^+ t}
p_{kl}(\theta,\theta'),\hat\Psi_r(\theta)\otimes
\ov{\hat\Psi_s}(\theta')\rangle
\nonumber\\
&=&o(1),\,\,\,\,t\to\infty.
\eeqn
The same decay as $t\to+\infty$ is valid if
we substitute
$\omega^-_{kl}$
in  (\ref{6.13})
instead of $\omega^+_{kl}$
for all $k,l\in {\ov n}$ except when
$k,l\in(r_{\sigma-1},r_\sigma]$.
For  $k,l\in(r_{\sigma-1},r_\sigma]$
we have $\om_k(\theta)\equiv\om_l(\theta)$.
Hence, 
by the  arguments of type  (\ref{6.91})-(\ref{6.16'})
and (\ref{6.93}), we obtain 
\beqn\la{6.14}
J^{-}_{kl}(t)\!\!&=&\!\!(2\pi)^{-2d}
\langle
g_m g_{m'}
\de(\bar\theta\!-\!\bar\theta')(2\pi)^{d-1}i
\PV\frac{1}{
\tg(\theta_d\!-\!\theta'_d)/2}
\,e^{- i\omega_{kl}^- t}
p_{kl}(\theta,\theta'),
\hat \Psi_r(\theta)\otimes
 \ov{\hat\Psi_s}(\theta')\rangle\nonumber\\
\!\!&=&\!\!(2\pi)^{-d}\int\limits_{T^{d}}
g_m g_{m'}\chi_{kl}\,
p_{kl}(\theta,\theta)
\sgn\Big( \frac{\partial \om_k}{\partial
\theta_d}(\theta)\Big)
\ov{\hat\Psi_r}(\theta)\hat\Psi_s(\theta)\,d\theta
+o(1),\,\,\,t\to+\infty.\,\,\,\,\,\,\,\,\,\,\,\,
\eeqn
 From (\ref{6.92}), (\ref{6.13})
 it follows 
that  for any $k,l\in {\ov n}$ as $t\to\infty$,
\be\la{6.15}
\langle
g_m g_{m'}\de(\bar\theta\!-\!\bar\theta')(2\pi)^{d-1}i
\PV\frac{1}{\tg(\theta_d\!-\!\theta'_d)/2}
\cos(\omega_{kl}^\pm t)
p_{kl}(\theta,\theta'),
\hat \Psi_r(\theta)\otimes
\ov{\hat\Psi_s}(\theta')\rangle=o(1),
\ee
since
 the signes in (\ref{6.93}) and (\ref{6.14})
are oppposite. 
Similarly, by  (\ref{6.92}) and (\ref{6.13}), we have
\be\la{6.18}
\langle
g_m g_{m'}\de(\bar\theta\!-\!\bar\theta')(2\pi)^{d-1}i
\PV\frac{1}{\tg(\theta_d\!-\!\theta'_d)/2}
\sin{\big(\omega_{kl}^+ t\big)}
p_{kl}(\theta,\theta'),
\hat \Psi_r(\theta)\otimes
\ov{\hat\Psi_s}(\theta')\rangle=o(1),\,\,\,\,t\to\infty.
\ee
The same relation holds if we substitute
$\omega_{kl}^-$
in the LHS of (\ref{6.18}) 
instead of  $\omega_{kl}^+$
for all $k,l\in {\ov n}$
except when $k,l\in(r_{\sigma-1},r_\sigma]$.
At last, 
using (\ref{6.14}), we get:
\beqn\la{6.17}
&&(2\pi)^{-2d}
\langle
g_m g_{m'}\de(\bar\theta\!-\!\bar\theta')(2\pi)^{d-1}i
\PV\frac{1}{\tg(\theta_d-\theta'_d)/2}
\sin({\omega_{kl}^- t)}
~p_{kl}(\theta,\theta'),
\hat \Psi_r(\theta)\otimes
\ov{\hat\Psi_s}(\theta')\rangle\nonumber\\
&&=(2\pi)^{-d}
\langle
g_m g_{m'}\chi_{kl}\,
 i\sgn\Big(\frac{\pa\om_k}{\pa\theta_d}(\theta)\Big)p_{kl}(\theta,\theta),
\hat\Psi_r(\theta)\otimes
\ov{\hat\Psi_s}(\theta)\rangle+o(1),\,\,\,t\to+\infty.\,\,\,\,\,\,\,\,\,\,\,\,\,
\eeqn
Here (see {\it Steps 1, 2}) by $p_{kl}(\theta,\theta)$
we denote 
$$
p_{kl}(\theta,\theta)\equiv
B_{rk}(\theta)\Big(M_2^-(\theta,\theta)\Big)_{kl}
B^*_{ks}(\theta)\equiv
B_{rk}(\theta)\Big(
B^*(\theta)M_0^-(\theta)B(\theta)\Big)_{kl}
B^*_{ks}(\theta),
$$
where $M_0^-(\theta)$ is defined by (\ref{Pi-}).

{\it Step 4}\,
Now we  return to the  RHS of (\ref {6.20}).
Let us substitute (\ref{6.90}) in   (\ref{6.20}).
Then by (\ref{6.15})
the summands  in the RHS (\ref{6.20})
with $\cos \om_{kl}^\pm t$ tend to zero.
Further, by (\ref{6.18}), (\ref{6.17})
 only integrals with 
$\sin \om_{kl}^- t$, 
$k,l\in(r_{\sigma-1},r_\sigma]$,
$\sigma=1,\dots,s+1$ contribute to a limit.
Finally, (\ref{6.20}), (\ref{6.90}) 
and (\ref{6.13})-(\ref{6.17})
imply,
\beqn
&&\langle Q^-_{t}(x,y),\Psi(x)\otimes
\Psi(y)\rangle
\nonumber\\
\!&=&\!\!\!
(2\pi)^{-d}\sum\limits_{m,m'}
\langle g_m g_{m'} 
B(\theta)\Big[\chi_{kl}
i\sgn\Big(
\frac{\pa\om_k}{\pa\theta_d}(\theta)
\Big)
M_2^-(\theta,\theta)_{kl}\Big]_{k,l\in {\ov n}}
B^*(\theta),\hat\Psi(\theta)\otimes
\ov{\hat\Psi}(\theta)\rangle
\!+\!o(1)
\nonumber\\
\!&=&\!\!\!(2\pi)^{-d}
\sum\limits_{m,m'}
\langle g_m g_{m'}\,
\hat q^-_{\infty}(\theta),\hat\Psi(\theta)\otimes
\ov{\hat\Psi}(\theta)\rangle
+o(1)
\nonumber\\
\!&=&\!\!\!
\langle
 q^-_{\infty}(x-y),\Psi(x)\otimes\Psi(y)\rangle
+o(1),
\nonumber
\,\,\,\,t\to+\infty.\,\,\,\,\,\,\,\,\,\,\,\,\,\,\,\,\,\,\,\,\,\,\,\,\,\,\,\,\,\,\,\,\,\,\,\,\,\,\,\,\,\,\,\,\,\,\,\,\,\,\,\,\,\,\,\,\,\,\,\,\,\,\,\,\,\,\,\,\,\,\,\,\Box
\eeqn

\subsection{ Convergence of $Q^r_t(x,y)$}

\begin{lemma}\la{Qt4}
$\lim\limits_{t\to\infty}
\langle Q^r_{t}(x,y),\Psi(x)\otimes
\Psi(y))\rangle=0$ 
for any $\Psi\in{\cal D}^0$.
\end{lemma}
{\bf Proof.}\,
 {\it Step 1}\,
We develop the method \ci[p.140]{BPT}.
Let us define  (as in (\ref{YP}))
$$
\Phi(x',t):=
\sum\limits_{x\in\Z^d}{\cal G}^T_t(x-x')\Psi(x).
$$
Then using (\ref{Qta}) we have,
\beqn\la{5.100}
\langle Q^r_t(x,y),\Psi(x)\otimes\Psi(y)\rangle=
\sum\limits_{x'\in \Z^d}
\sum\limits_{y'\in \Z^d}
Q^r(x',y')\Phi(x',t)\Phi(y',t)
=\sum\limits_{z'\in \Z^d}
{\cal F}_t(z'),
\eeqn
where
\be\la{calF}
{\cal F}_t(z'):=
\sum\limits_{y'\in \Z^d}
Q^r(y'+z',y')
\Phi(y'\!+\!z',t)\Phi(y',t).
\ee
The estimates
(\ref{4.9'}), (\ref{4.9})
and definition (\ref{d1'''})
imply the same estimate for
$Q^r(x,y)$:\\
$|Q^r(x,y)| \le Ce_0\varphi^{1/2}(|x-y|)$.
Hence,
 the Cauchy-Schwartz inequality and 
 (\ref{esPhi})   imply
\beqn
|{\cal F}_t(z')|&\le&
\sum\limits_{y'\in \Z^d}
\Vert Q^r(y'+z',y')\Vert\,
|\Phi(y'+z',t)|\,|\Phi(y',t)|
\nonumber\\
&\le&
C\varphi^{1/2}(|z'|)
\sum\limits_{y'\in \Z^d}
|\Phi(y'+z',t)|\,|\Phi(y',t)|\le
\!C_1\varphi^{1/2}(|z'|)
\Vert \Psi\Vert^2_{V},
\eeqn
where $\Vert \Psi\Vert^2_{V}$
is defined by (\ref{nDV}).
 Hence, (\ref{1.12}) and condition {\bf E6}
 imply
\be\la{5.101}
\sum\limits_{z'\in \Z^d}
|{\cal F}_t(z')|\le
C(\Psi)
\sum\limits_{z'\in \Z^d}\varphi^{1/2}(|z'|)\le
C_1<\infty,
\ee
and the series (\ref{5.100})
 converges uniformly in $t$.
Therefore, it suffices  to prove that
\be\la{convF}
\lim_{t\to\infty}{\cal F}_t(z')=0
\,\,\,\,\mbox{for each } z'\in \Z^d.
\ee
{\it Step 2}\,
Let us prove (\ref{convF}).
Condition {\bf S1} and (\ref{d1'''}) imply that
$
Q^r(y'+z',y')=q^r(\bar z',y'_d+z'_d,y'_d),
$
 where
\be\la{3b}
\lim_{y'_d\to\pm\infty}
q^r(\bar z',y'_d+z'_d,y'_d)= 0,\,\,\,
\mbox{for }\,(\ov z',z'_d)\in\Z^d.
\ee
Hence,
$\forall\ve>0$ there exists $N\in\N$
so large that $|q^r(\bar z',y'_d\!+\!z'_d,y'_d)|<\ve$
for $|y'_d|>N$.
Respectively,
 decompose the series (\ref{calF})
into two series:
${\cal F}_t(z')=\sum\limits_{\bar y'\in \Z^{d-1}}
\sum\limits_{|y'_d|>N}\ldots+
\sum\limits_{\bar y'\in \Z^{d-1}}
\sum\limits_{|y'_d|<N}\ldots$.
By (\ref{esPhi}) and condition {\bf E6},
 the first series  is estimated by
\beqn\la{7.31}
\Big|\sum\limits_{\bar y'\in \Z^{d-1}}
\sum\limits_{|y'_d|>N}
q^r(\bar z',y'_d+z'_d, y'_d)
\Phi(y'\!+\!z',t)\Phi(y',t)\Big|\le
\ve \sum\limits_{y'\in\Z^d}
|\Phi(y',t)|^2
\le
\ve\, C(\Psi).
\eeqn
Note that
$q^r(\bar z',y'_d+z'_d, y'_d)$
does not depend on $\ov y'$.
Then we can rewrite the second series  by the
Parseval identity as
\beqn\la{2sum}
&&\sum\limits_{|y'_d|<N}
q^r(\bar z',y'_d+z'_d, y'_d)
\sum\limits_{\bar y'\in \Z^{d-1}}
\Phi(y'\!+\!z',t)\Phi(y',t)
\nonumber\\
&=&
(2\pi)^{-2d+2}
\sum\limits_{|y'_d|<N}
q^r(\bar z',y'_d\!+\!z'_d, y'_d)
\int\limits_{T^{d-1}}
F_{\bar y'\to\bar \theta}[\Phi(y'+z',t)]\,
\ov{F_{\bar y'\to\bar \theta}[\Phi(y',t)]}\,d\bar\theta.
\,\,\,\,\,\,\,
\eeqn
It remains to prove that the 
 integral in the RHS of (\ref{2sum})
tends to zero as $t\to\infty$ for fixed 
$z'\in\Z^d$ and  $|y'_d|<N$.
First,
let us note that for
the integrand in (\ref{2sum}) 
the following uniform bound holds,
\be\la{7.34}
|F_{\bar y'\to\bar \theta}[\Phi(y'+z',t)]
\ov{F_{\bar y'\to\bar \theta}[\Phi(y',t)]}|\le 
G(\bar\theta),\,\,\,\,t\ge0,\,\,\,
\mbox{ where }\, G(\bar \theta)\in L^1(T^{d-1}).
\ee
Indeed, rewrite  the function 
$F_{\bar y'\to\bar \theta}[\Phi(y',t)]$
in the form
\be\la{7.33}
F_{\bar y'\to\bar \theta}[\Phi(y',t)]
=(2\pi)^{-1}\int\limits_{T^1}
e^{-i\theta_d y'_d}\,\hat\Phi(\theta,t)\,d\theta_d
=
(2\pi)^{-1}\int\limits_{T^1}
e^{-i\theta_d y'_d}\,\hat{\cal G}^*_t(\theta)
\hat\Psi(\theta)\,d\theta_d.
\ee
Therefore, 
\beqn
|F_{\bar y'\to\bar \theta}[\Phi(y'\!+\!z',t)]
\ov{F_{\bar y'\to\bar \theta}[\Phi(y',t)]}|
\!&\le&\!
C\Big(\int\limits_{T^1}
\Vert \hat{\cal G}^*_t(\theta)\Vert\,
|\hat\Psi(\theta)|\,d\theta_d\Big)^2\le
C_1\int\limits_{T^1}
\Vert \hat{\cal G}^*_t(\theta)\Vert^2\,
|\hat\Psi(\theta)|^2\,d\theta_d
\nonumber\\
&\le& C_2\int\limits_{T^1}
\Vert(1+\Vert\hat V^{-1}(\theta)\Vert )
|\hat\Psi(\theta)|^2\,d\theta_d:=G(\bar\theta)
\eeqn
and (\ref{7.34}) follows from condition
{\bf E6}.
Therefore,
 it suffices to prove
that the integrand in the RHS of (\ref{2sum})
tends to zero  as $t \to\infty$  for a.a. fixed
 $\bar \theta\in T^{d-1}$.
We use the  finite partition of unity
(\ref{part}) (remember that
$\Psi\in{\cal D}^0$) 
and split the function
$F_{\bar y'\to\bar \theta}[\Phi(y',t)]$
 into the sum of the integrals:
\be\la{7.37}
F_{\bar y'\to\bar \theta}[\Phi(y',t)]=
\sum\limits_m\sum\limits_{\pm,k\in\ov n}\,
\int\limits_{T^1}
g_m(\theta) e^{-i\theta_d y'_d}e^{\pm i\om_k(\theta)t}
a^\pm_k(\theta)\hat \Psi(\theta)\,d\theta_d,
\,\,\,\,\Psi\in{\cal D}^0.
\ee
The eigenvalues $\om_k(\theta)$ 
and the matrices $a^\pm_k(\theta)$  
are 
real-analytic functions inside 
the $\supp g_m$ for every $m$.
From Definition \ref{dC} i)
and conditions~{\bf  E4, E6}
it follows that
 mes$\{\theta_d\in T^1:\,\nabla_{\theta_d}\om_k(\theta)=0\}=0$
for a.a. fixed $\ov \theta\in T^{d-1}$.
Hence,
the integrals  in (\ref{7.37})
vanish as $t\to\infty$
by  the Lebesgue-Riemann theorem.
 \hfill$\Box$ \\

Finally, Lemmas \ref{Qt1}, \ref{Qt3} and \ref{Qt4} imply
the convergence (\ref{4.4}) for $\Psi\in{\cal D}^0$.
Then   (\ref{4.4})
follows
 for any $\Psi\in{\cal D}$ by Lemma \ref{lcc}
(see section 6.1).
Proposition \ref{p4.1} is proved.
\hfill$\Box$

\setcounter{equation}{0}

\section{Bernstein's argument}

  \subsection{
Oscillatory integrals  and stationary phase method}

To prove  (\ref{2.6i})
we evaluate $\langle Y(\cdot,t),\Psi\rangle $
by (\ref{YP}), where
\be\la{frep'}
\Phi(x,t):=F_{\theta\to x}^{-1}[
\hat{\cal G}^*_t(\theta)\hat \Psi(\theta)]=(2\pi)^{-d}
\int\limits_{T^d} e^{-i\theta x}
\hat{\cal G}^*_t( \theta)\hat\Psi( \theta)~d \theta,
\,\,\,x\in\Z^d.
\ee
Similarly to (\ref{7.37}) or (\ref{qtpar}) 
using the partition of unity (\ref{part}) we get
\be\la{frepe}
\Phi(x,t)=
\sum\limits_{m}
\sum\limits_{\pm,\,k\in \ov n }~~
\int\limits_{T^d}
g_m(\theta) e^{-i(\theta x\pm\om_k t)}
a^\pm_k(\theta)
\hat\Psi( \theta)~d \theta,
\,\,\,\,\Psi\in{\cal D}^0,
\ee
where $\om_k(\theta)$
and $a^\pm_k(\theta)$
are real-analytic functions inside the $\supp g_m$ for every $m$.

Note that  $\Phi(t):=\Phi(\cdot,t)$ is the solution to the
"conjugate"
 equation
(cf (\ref{CP}), (\ref{A}))
\be\la{CPA}
\dot \Phi(t)={\cal A}'\Phi(t),\,\,\,t\in\R;
\quad\quad
{\cal A}'=
\left(
 \begin{array}{cc}
0 & - {\cal V}\\
1 & 0
\end{array}\right),
\ee
which is obvious in the Fourier transform.
Therefore, the solutions $Y(t)=(Y^0(t),Y^1(t))$ and
$\Phi(t)= (\Phi^0(t),\Phi^1(t))$
to the equations (\ref{CP}) and (\ref{CPA})
coincide up to order of the components.
Hence, $\Phi(x,t)$ has corresponding  dispersive properties.

We will deduce (\ref{2.6i}) by analyzing the
propagation
of the solution $\Phi(x,t)$ to Eqn
(\ref{CPA}), in different
directions $x=vt$ with $v\in\R^d$.
For this purpose, we apply the stationary phase method
 to the oscillatory integral (\ref{frepe})
along the rays $x=vt$, $t>0$. Then the phase  becomes
$(\theta v\pm\om_k(\theta))t$, and its stationary
 points are the solutions to
the equations $v=\mp\nabla\om_k(\theta)$.

Recall that we can restrict ourselves by $\Psi\in{\cal D}^0$,
hence $\hat\Psi(\theta)=0$ in the points $\theta\in T^d$
with  degenerate Hessian $D_k(\theta)$ (see {\bf E4}).
Therefore, the stationary phase method leads to the following
 two different types  of the
asymptotic behavior of $\Phi(vt,t)$ as $t\to\infty$:
\smallskip\\
{\bf I.} For the velocity $v$ inside the light cone:
$v=\pm\nabla\om_k(\theta)$,
$\theta\in T^d \setminus {\cal C}$.
Then
\be\la{spd}
 \Phi(vt,t)={\cal O}( t^{-d/2}).
\ee
{\bf II.} For the velocity $v$ outside the light cone:
$v\ne\pm\na\om_k(\theta)$,
$\theta\in T^d\setminus{\cal C}$, $k\in \ov n$.
Then
\be\label{rd}
\Phi(vt,t)={\cal O}( t^{-p}),\,\,\,\, \forall p>0.
\ee
The asymptotics of the types {\bf I} and {\bf II}  allow us to
incorporate
the Bernstein-type approach
developed  in \cite{BPT} for case $d=1$  and
in \cite{DKKS,DKRS} for  continuous
 Klein-Gordon and wave equations for $d\ge 1$.
We formalize (\ref{spd}), (\ref{rd}) as follows.
\begin{lemma}\la{l5.3}
For any fixed $\Psi \in {\cal D}^0$  the following
bounds hold:
\be\la{bphi}
\!\!\! i)~~~~~
\sup_{x\in\Z^d}|\Phi(x,t)| \le  C~t^{-d/2}.
~~~~~~~~~~~~~~~~~~~~~~~~~~~~~~~~~~~~~~~~~~~~~~~~~~~~~~~~~~~~~~~~~~~~~~~~~~~~
\ee
ii) For  any $p>0$ there exist   
$C_p,\ga>0$ s.t. 
\be\la{conp}
|\Phi(x,t)|\le C_p(1+|x|+|t|)^{-p},\quad\quad
 |x|\ge\gamma t.
\ee
\end{lemma}
{\bf Proof}
Consider $\Phi(x,t)$ along each ray   
$x=vt$ with  arbitrary $v\in\R^d$. 
Substituting  to  
(\ref{frepe}), we get 
\be\la{freper}   
\Phi(vt,t)= \sum_{m}  
\sum\limits_{\pm,\,\,k\in \ov n }~~   
\int\limits_{T^d} g_m(\theta)
e^{-i(\theta v\pm\om_k(\theta)) t}   
a^\pm_k(\theta)   
\hat\Psi( \theta)~d \theta,\,\,\,\,\hat\Psi\in{\cal D}^0.   
\ee   
This is a sum of oscillatory integrals with the phase   
functions $\phi_k^\pm(\theta)=   
\theta v\pm\om_k(\theta)$ 
and the amplitudes $a^\pm_k(\theta) $  
which are real-analytic 
functions of the $\theta$   
inside 
the $\supp g_m$. 
Since $\om_k(\theta)$ is real-analytic, 
 each function $\phi_k^\pm$   
 has no more than a   
 finite number  of  stationary points $\theta\in\supp g_m$,   
solutions to the equation $v=\mp\nabla\om_k(\theta)$.   
The stationary points are non-degenerate for 
$\theta\in\supp g_m$   
by  (\ref{part}), 
Definition \ref{dC} and ${\bf E4}$  
since   
\be\la{Hess}   
{\rm det}\Big(\frac{\pa^2 \phi_k^\pm}
{\pa \theta_i\pa \theta_j}\Big)=   
\pm D_k(\theta)\not= 0,\,\,\,\,\,\theta\in \supp g_m.   
\ee   
At last, $\hat\Psi( \theta)$ is smooth since  $\Psi\in{\cal D}$. 
Therefore, $\Phi(vt,t)={\cal O}(t^{-d/2})$   
 according to the standard stationary phase method   
 \ci{F, RS3}.  This implies the bounds (\ref{bphi}) 
in each cone $|x|\le ct$ with any finite $c$. 
 
 Further, denote by   
$\bar v:=\max_m\max_{k\in\overline n} 
\max\limits_{\theta\in \supp g_m}  
|\nabla \om_k(\theta)|.   
$   
Then for $|v|>\bar v$ 
the stationary points do not exist on the $\supp \hat\Psi $. 
Hence, the integration by parts as in \ci{RS3} yields 
$\Phi(vt,t)={\cal O}(t^{-p})$ for any $p>0$. 
On the other hand, the integration by parts in  
(\ref{frepe})  
implies similar bound  
$\Phi(x,t)={\cal O}\Big(\ds(t/|x|)^l\Big)$ for any $l>0$. 
Therefore, (\ref{conp}) follows with any 
$\ga>\ov v$.  
Now the bounds (\ref{bphi}) follow everywhere. 
\hfill$\Box$
\subsection{`Rooms - corridors' partition}
The remaining constructions in the proof of (\ref{2.6i}) are
 similar to \ci{DKKS, DKS1}. 
However, the proofs are not identical since here we consider
non translation-invariant case.


Let us 
introduce a `room-corridor'  partition of the
ball $\{x\in\Z^d:~|x|\le \ga t\}$,
 with $\ga$ from (\ref{conp}).
For $t>0$ we choose
$ \De_t$ and $\rho_t\in\N$.
Asymptotical relations between $t$, $\De_t$ and  $\rho_t$
are specified below.
Let us set $h_t=\De_t+\rho_t$ and   
\be\la{rom}   
a^j=jh_t,\,\,\,b^j=a^j+\De_t,\,\,\,   
j\in\Z,\,\,\,\,\,\,N_t=[(\gamma t)/h_t].   
\ee   
We call the slabs $R_t^j=\{x\in\Z^d:
 |x|\le N_t h_t,\,a^j\le x_d< b^j\}$   
the `rooms',   
$C_t^j=\{x\in\Z^d: |x|\le  
N_t h_t,\, b^j\le x_d<  a^{j+1}\}$  the `corridors'   
and $L_t=\{x\in\Z^d: |x|> N_t h_t\}$ the 'tails'.   
Here  $x=(x_1,\dots,x_d)$,   
$\De_t$ is the width of a room, and   
$\rho_t$  of a corridor. 
Let us denote  by   
 $\chi_t^j$ the indicator of the room $R_t^j$, 
 $\xi_t^j$ that of the corridor $C_t^j$, and  
$\eta_t$ that of the tail $L_t$. 
Then  
\be\la{partB}   
{\sum}_t   
[\chi_t^j(x)+\xi_t^j(x)]+ \eta_t(x)=1,\,\,\,x\in\Z^d,  
\ee   
where the sum ${\sum}_t$ stands for   
$\sum\limits_{j=-N_t}^{N_t-1}$.   
Hence, we get the following  Bernstein's type representation:   
\be\la{res}   
\langle Y_0,\Phi(\cdot,t)\rangle = {\sum}_t   
[\langle Y_0,\chi_t^j\Phi(\cdot,t)\rangle +   
\langle Y_0,\xi_t^j\Phi(\cdot,t)\rangle ]+   
\langle Y_0,\eta_t\Phi(\cdot,t)\rangle.   
\ee   
Let us define  the   
random variables   
 $ r_{t}^j$, $ c_{t}^j$, $l_{t}$ by   
\be\la{100}   
r_{t}^j= \langle Y_0,\chi_t^j\Phi(\cdot,t)\rangle,~~   
c_{t}^j= \langle Y_0,\xi_t^j\Phi(\cdot,t)\rangle,   
\,\,\,l_{t}= \langle Y_0,\eta_t\Phi(\cdot,t)\rangle.   
\ee   
Then  (\ref{res}) becomes 
\be\la{razli}   
\langle Y_0,\Phi(\cdot,t)\rangle =   
{\sum}_t   
(r_{t}^j+c_{t}^j)+l_{t}.   
\ee   

\begin{lemma}  \la{l5.1}
    Let {\bf S0--S3} hold and $\Psi\in{\cal  D}^0$.
The following bounds hold for $t>1$:
\beqn
E|r^j_{t}|^2&\le&  C(\Psi)~\De_t/ t,\,\,\,\forall j,\la{106}\\
E|c^j_{t}|^2&\le& C(\Psi)~\rho_t/ t,\,\,\,\forall j,\la{106''}\\
E|l_{t}|^2&\le& C_p(\Psi)~t^{-p},\,\,\,\,\forall p>0.
\la{106'''}
\eeqn
\end{lemma}
{\bf Proof}\, (\ref{106'''})
follows from (\ref{conp}) and
Proposition \ref{l4.1}, i).
We discuss  (\ref{106}) only,  (\ref{106''})
is done in a similar way.
Let us express $E|r_t^j|^2$  in the correlation matrices.
Definition (\ref{100})   implies 
 \be\la{100rq}
E|r_{t}^j|^2= \langle Q_0(x,y),
 \chi_t^j(x)\Phi(x,t)\otimes\chi_t^j(y)\Phi(y,t)\rangle.
\ee
 According to
(\ref{bphi}),  Eqn (\ref{100rq})
 implies that
\beqn\la{er}
E|r_{t}^j|^2&\le&
C t^{-d}
\sum\limits_{x,y}
\chi_t^j(x)\Vert Q_0(x,y)\Vert \nonumber\\
&=&Ct^{-d}\sum\limits_{x}
\chi_t^j(x)
\sum\limits_{y\in\Z^d}
\Vert Q_0(x,y)\Vert\le C \De_t/t,
\eeqn
where $\Vert Q_0(x,y)\Vert $ stands for the norm of a matrix
$\left(Q_0^{ij}(x,y)\right)$.
Therefore,  (\ref{er})   follows
by Proposition \ref{l4.1}, i).
\hfill$\Box$\\
\medskip

Now we prove the convergence  (\ref{2.6i}).
As was said, we use a version
of the Central Limit Theorem
developed by Ibragimov and Linnik.
If  ${\cal Q}_{\infty}(\Psi,\Psi)=0$,
 the convergence (\ref{2.6i}) is obvious.
In fact, then,
\beqn
&&| E\exp\{i \langle Y_0,\Phi(\cdot,t)\rangle\} -
 \hat\mu_{\infty}(\Psi)|
= E|\exp\{i \langle Y_0,\Phi(\cdot,t)\rangle\} -
 1|\le
E|\langle Y_0,\Phi(\cdot,t)\rangle|
\nonumber\\
&\le&
\Big(E|\langle Y_0,\Phi(\cdot,t)\rangle|^2
\Big)^{1/2}
=\Big(\langle Q_0(x,y),\Phi(x,t)\otimes
\Phi(y,t)\rangle
\Big)^{1/2}=
\Big({\cal Q}_t(\Psi,\Psi)\Big)^{1/2},
\eeqn
where ${\cal Q}_t(\Psi,\Psi)\to 
{\cal Q}_\infty(\Psi,\Psi)=0$, $t\to\infty$.
Therefore,  (\ref{2.6i})
follows from (\ref{4.4}).
Thus, we may assume that for a given $\Psi\in{\cal D}^0$,
\be\la{5.*}
{\cal Q}_{\infty}(\Psi,\Psi)\not=0.
\ee
Let us choose  $0<\de<1$ and
\be\la{rN}
\rho_t\sim t^{1-\delta},
~~~\De_t\sim\fr t{\log t},~~~~\,\,\,t\to\infty.
\ee
\begin{lemma}\la{r}
The following limit holds true:
\be\la{7.15'}
N_t\Bigl(
\varphi(\rho_t)+\Bigl(
\frac{\rho_t}{t}\Bigr)^{1/2}\Bigr)
+
N_t^2\Bigl(
\varphi^{1/2}(\rho_t)+\frac{\rho_t}{t}\Bigr)
\to 0 ,\quad t\to\infty.
\ee
\end{lemma}
{\bf Proof}.
Function $\varphi(r)$ is nonincreasing, hence  by
(\ref{1.12}),
\be\la{1111}
r^{d}\varphi^{1/2}(r)=
d\int\limits_0^r
s^{d-1}\varphi^{1/2}(r)\,ds\le
d \int\limits_0^r
s^{d-1}\varphi^{1/2}(s) \,ds\le C\ov\varphi<\infty.
\ee
Furthermore, (\ref{rN}) implies
that
$h_t=\rho_t+\De_t\sim \ds\frac{t}{\log t}$, $t\to\infty$.
Therefore, $N_t\sim\ds\frac{t}{h_t}\sim\log t$.
Then  (\ref{7.15'}) follows by (\ref{1111})
and (\ref{rN}).
\hfill$\Box$\\

By the triangle inequality,
\beqn
|E\exp\{i \langle Y_0,\Phi(\cdot,t)\rangle  \}
-\hat\mu_{\infty}(\Psi)|&\le &
|E\exp\{i \langle Y_0,\Phi(\cdot,t)\rangle  \}-
E\exp\{i{{\sum}}_t r_{t}^j\}|
\nonumber\\
&&+|\exp\{-\frac{1}{2}{\sum}_t E|r_{t}^j|^2\} \!-\!
\exp\{-\frac{1}{2} {\cal Q}_{\infty}(\Psi, \Psi)\}|
 \nonumber\\
&&+ |E \exp\{i{\sum}_t r_{t}^j\} \!-\!
\exp\{-\frac{1}{2}{\sum}_t E|r_{t}^j|^2\}|\nonumber\\
&\equiv& I_1+I_2+I_3. \la{4.99}
\eeqn
We are going to   show  that all summands
$I_1$, $I_2$, $I_3$  tend to zero
 as  $t\to\infty$.\\
{\it Step (i)}
Eqn (\ref{razli}) implies
\beqn\la{101}
I_1&=&|E\exp\{i{\sum}_t r^j_{t} \}
\Big(\exp\{i{\sum}_t c^j_{t}+il_{t}\}-1\Big)|\nonumber\\
&\le&
 {\sum}_t E|c^j_t|+E|l_{t}|\le
{\sum}_t(E|c^j_t|^2)^{1/2}+(E|l_{t}|^2)^{1/2}.
\eeqn
From (\ref{101}), (\ref{106''}), (\ref{106'''})
  and (\ref{7.15'}) we obtain that
\be\la{103}
I_1\le C_p t^{-p}+ C N_t(\rho_t/t)^{1/2}\to 0,~~t\to \infty.
\ee
{\it Step (ii)}
By the triangle inequality,
\beqn
I_2&\le& \frac{1}{2}
|{\sum}_t E|r_t^j|^2-
 {\cal Q}_{\infty}(\Psi, \Psi) |
\le
 \frac{1}{2}\,
|{\cal Q}_{t}(\Psi, \Psi)-{\cal Q}_{\infty}(\Psi, \Psi)|
\nonumber\\
&&+ \frac{1}{2}\, |E\Bigl({\sum}_t r_t^j\Bigr)^2
-{\sum}_tE|r_t^j|^2| +
 \frac{1}{2}\, |E\Bigl({\sum}_t r_t^j\Bigr)^2
-{\cal Q}_{t}(\Psi, \Psi)|\nonumber\\
&\equiv& I_{21} +I_{22}+I_{23}\la{104},
\eeqn
where ${\cal Q}_{t}$ is  a quadratic form with
the  matrix kernel $\Big(Q_t^{ij}(x,y)\Big)$.
(\ref{4.4}) implies that
 $I_{21}\to 0$.
As to  $I_{22}$,  we first have that
\be\la{i22}
I_{22}\le \sum\limits_{j< l}
| E r_t^j r_t^l|.
\ee
The next lemma   is a corollary of  \ci[Lemma 17.2.3]{IL}.
\begin {lemma}\la{il}
  Let ${\cal A},{\cal B}$ be the subsets of $\Z^d$
with the distance {\rm dist}$({\cal A}, {\cal B})\ge r>0$, and 
$ \xi, \eta$ be random variables  
on the probability space 
$({\cal H}_\al,{\cal B}({\cal H}_\al),\mu_0)$.
Moreover,  let $ \xi$ be 
 measurable with respect to the 
$\sigma$-algebra $\sigma({\cal A})$,
  $\eta$  with respect to the
$\sigma$-algebra $\sigma({\cal B})$. 
Then\\  
i)\, 
$\hspace{.5mm}   
|E\xi\eta-E\xi E\eta|\le  
C ab~  
\varphi^{1/2}(r)  
$
if $(E|\xi|^2)^{1/2}\le a$ and $(E|\eta|^2)^{1/2}\le b$.
\\  
ii) 
$  
|E\xi\eta-E\xi E\eta|\le Cab~  
\varphi(r)  \hspace{.5mm}
$  
\,\,\,\,\,\,if $|\xi|\le a$ and $|\eta|\le b$ a.e.
\end{lemma}
We apply Lemma \re{il} to deduce that
$I_{22}\to 0$ as $t\to\infty$.
Note  that
$r_t^j=
\langle Y_0,\chi_t^j \Phi(\cdot,t)\rangle $
is measurable
with respect to the $\sigma$-algebra  $\sigma(R_t^j)$.
The distance
between the different rooms $R_t^j$
is greater or equal to
$\rho_t$ according to
 (\ref{rom}).
Then (\ref{i22}) and {\bf S1}, {\bf S3} imply,
together with
Lemma \ref{il} i) and (\ref{106}), that
\be\la{i222}
I_{22}\le
C N_t^2\varphi^{1/2}(\rho_t),
\ee
which vanishes as $t\to\infty$ because of
 (\ref{7.15'}).
Finally, it remains to check
 that $I_{23}\to 0$,
$t\to\infty$.
We have
$$
{\cal Q}_t(\Psi,\Psi)
=E\langle Y_0,\Phi(\cdot,t)\rangle^2
=E\Big({\sum}_t (r_t^j+c_t^j)+l_t\Big)^2,
$$
according  to (\ref{razli}).
Therefore, by the
Cauchy-Schwartz inequality,
\beqn
I_{23}&\le&
 |E\Bigl({\sum}_t r_t^j\Bigr)^2
- E\Bigl({\sum}_t r_t^j +
{\sum}_t c_t^j+l_t\Bigr)^2 |\nonumber\\
& \le&
C N_t{\sum}_t E |c_t^j|^2  +
C_1\Bigl(
E({\sum}_t r_t^j)^2\Bigr)^{1/2}
\Bigl(
N_t{\sum}_t E|c_t^j|^2+E |l_t|^2\Bigr)^{1/2}
+C  E |l_t|^2.\la{107}
\eeqn
Then  (\ref{106}), (\ref{i22}) and (\ref{i222})
imply
$$
E({\sum}_t r_t^j)^2\le
{\sum}_tE|r_t^j|^2 +2{\sum}_{j<l}|E r_t^j r_t^l|
\le
CN_t\De_t/t+C_1N_t\varphi^{1/2}(\rho_t)\le C_2<\infty.
$$
Now
(\ref{106''}), (\ref{106'''}),
 (\ref{107}) and (\ref{7.15'}) yield
\be\la{106'}
I_{23}\le C_1  N_t^2\rho_t/t+C_2 N_t(\rho_t/t)^{1/2}
+C_3 t^{-p} \to 0,~~t\to \infty.
\ee
So,  all terms $I_{21}$, $I_{22}$, $I_{23}$
in  (\ref{104})
tend to zero.
Then
 (\ref{104}) implies that
\be\la{108}
I_2\le
\frac{1}{2}\,
|{\sum}_{t}E|r_t^j|^2-
 {\cal Q}_{\infty}(\Psi, \Psi)|
\to 0,~~t\to\infty.
\ee
{\it Step (iii)}
It remains to verify that
$$I_3=|
E\exp\{i{\sum}_t r_t^j\}
-\exp\{-\fr12 {\sum}_t E|r_t^j|^2\}| \to 0,~~t\to\infty.
$$
 Lemma \ref{il} ii)
yields:
\beqn
&&|E\exp\{i{\sum}_t r_t^j\}-\prod\limits_{-N_t}^{N_t-1}
E\exp\{i r_t^j\}|
\nonumber\\
&\le&
|E\exp\{ir_t^{-N_t}\}\exp\{i\sum\limits_{-N_t+1}^{N_t-1} r_t^j\}  -
 E\exp\{ir_t^{-N_t}\}E\exp\{i\sum\limits_{-N_t+1}^{N_t-1} r_t^j\} |
\nonumber\\
&&+
|E\exp\{ir_t^{-N_t}\}E\exp\{i\sum\limits_{-N_t+1}^{N_t-1} r_t^j\}
-\prod\limits_{-N_t}^{N_t-1}
E\exp\{i r_t^j\}|
\nonumber\\
&\le& C\varphi(\rho_t)+
|E\exp\{i\sum\limits_{-N_t+1}^{N_t-1} r_t^j\}
-\prod\limits_{-N_t+1}^{N_t-1}
E\exp\{i r_t^j\}|.\nonumber
\eeqn
We then apply Lemma \ref{il}, ii) recursively
and get, according to Lemma \ref{r},
$$|E\exp\{i{\sum}_{t} r_t^j\}-\prod\limits_{-N_t}^{N_t-1}
E\exp\{i r_t^j\}|
\le
C N_t\varphi(\rho_t)\to 0,\quad t\to\infty.
$$
It remains to check that
$$|\prod\limits_{-N_t}^{N_t-1} E\exp\{ir_t^j\}
-\exp\{-\fr12{\sum}_{t} E|r_t^j|^2\}| \to 0,~~t\to\infty.
$$
According to the
standard statement of the
Central Limit Theorem
(see, e.g. \ci[Theorem 4.7]{P}),
it suffices to verify the  Lindeberg condition:
$\forall\de>0$,
$$
\frac{1}{\sigma_t}
{\sum}_t
 E_{\de\sqrt{\sigma_t}}
|r_t^j|^2 \to 0,~~t\to\infty.
$$
Here
$
\sigma_t\equiv {\sum}_t
E |r^j_t|^2,$
and $E_\ve f\equiv E (X_\ve f)$,
where $X_a$ is the indicator of the
event $|f|>\ve^2.$
Note that
(\ref{108})
and (\re{5.*}) imply  that
$\sigma_t \to
{\cal Q}_{\infty}(\Psi, \Psi)\not= 0,$
$t\to\infty.$
Hence it remains to verify that
\be\la{linc}
{\sum}_t
E_{\ve}
|r_t^j|^2 \to 0,~~t\to\infty,
 ~~ \mbox{ for any }\, \ve>0.
\ee
We check Eqn (\ref{linc}) in Section 9.
This will complete the proof of
Proposition \ref{l2.2}.
\hfill$\Box$

\setcounter{equation}{0}
\section{The Lindeberg condition}
The proof of (\ref{linc}) can be reduced to the case when
for some $\La\ge 0$ we have that
 \be\la{ass1}
  |u_0(x)|+|v_0(x)|\leq \La<\infty, ~~~x\in\Z^d.
  \ee
Then the proof of (\ref{linc}) is reduced  to
the convergence
\be\la{111}
{\sum}_t E|r_t^j|^4 \to 0,~~t\to\infty,
\ee
by using Chebyshev's inequality.
The general case can be covered by standard cutoff arguments
by
taking into account that
the bound (\ref{106}) for $E|r^j_t|^2$
depends only on  $e_0$ and $\varphi$.
 The last fact is obvious  from
(\ref{er}) and (\ref{qp}).
We deduce (\ref{111}) from
\begin{theorem}  \la{p5.1}
Let the conditions  of Theorem A  hold
and assume that (\ref{ass1}) is fulfilled.
Then
for any $\Psi\in {\cal D}^0$
the following bounds hold:
\be\la{112}
E|r_t^j|^4\le
C(\Psi) \La^4\De_t^2/t^2,   ~~t>1.
\ee
\end{theorem}
{\bf Proof.} {\it Step 1}~
Given four points $x^1,x^2,x^3,x^4\in\Z^d$, set: \\
$M_0^{(4)}(x^1,...,x^4)=E\left(Y_0(x^1)\otimes...\otimes Y_0(x^4)\right)$.
Then, similarly to (\ref{100rq}),
Eqns (\ref{ass1}) and (\ref{100}) imply 
\be\la{11.2}
E|r_t^j|^4=
\langle \chi_t^j(x^1) \ldots \chi_t^j(x^4)M_0^{(4)}(x^1,\dots,x^4),
\Phi(x^1,t)\otimes\dots\otimes\Phi(x^4,t)\rangle .
\ee
Let us analyze the domain of the  ${(\Z^d)^4}$ in
the RHS of (\ref{11.2}).
We partition  ${(\Z^d)^4}$ into three parts, $W_2$,  $W_3$  and $W_4$:
\be\la{Wi}
{(\Z^d)^4}=
\bigcup\limits_{i=2}^4 W_i,\quad
W_i=\{\bar x=(x^1,x^2,x^3,x^4)\in {(\Z^d)^4}:
|x^1-x^i|=\max\limits_{p=2,3,4}|x^1-x^p|\}.
\ee
Furthermore,  given
 $\bar x=(x^1,x^2,x^3,x^4)\in W_i$,
  divide $\Z^d$ into three parts
$S_j$, $j=1,2,3$: $\Z^d=S_1\cup S_2\cup S_3$,
by two
hyperplanes orthogonal to
the segment $[x^1,x^i]$ and  partitioning it into three equal segments,
where  $x^1\in S_1$ and $x^i\in S_3$.
Denote by
$x^p$, $x^q$  the  two remaining points
with $p,q\ne 1,i$.
Set:
 ${\cal A}_i=\{\bar x\in W_i:~
 x^p\in S_1,  x^q\in S_3 \}$,
 ${\cal B}_i=\{\bar x\in W_i:~
 x^p, x^q\not\in S_1 \}$ and
 ${\cal C}_i=\{\bar x\in W_i:~
 x^p, x^q\not\in S_3 \}$, $i=2,3,4$.
 Then
$W_i={\cal A}_i\cup {\cal B}_i\cup {\cal C}_i$.
Define the function ${\rm m}^{(4)}_0(\bar x)$, $\bar x\in{(\Z^d)^4},$
in the following way:
\beqn\la{M}
{\rm m}^{(4)}_0(\bar x)\Bigr|_{W_i}=
\left\{
\ba{ll}
M_0^{(4)}(\bar x)-
Q_0(x^1,x^p)\otimes
 Q_0(x^i,x^q),\quad \bar x\in{\cal A}_i,\\
M_0^{(4)}(\bar x),\quad \bar x\in{\cal B}_i\cup {\cal C}_i.
\ea
\right.
\eeqn
This determines ${\rm m}^{(4)}_0(\bar x)$
correctly  for  all quadruples  $\bar x$.
Note that
\beqn
\!&&\!\!\!\langle \chi_t^j(x^1) \ldots \chi_t^j(x^4)
Q_0(x^1,x^p)\otimes Q_0(x^i,x^q),
\Phi(x^1,t)\otimes\dots\otimes\Phi(x^4,t)\rangle
\nonumber\\
\!&&\!\!\!\!\!\!\!\!\!=\langle \chi_t^j(x^1)\chi_t^j(x^p) Q_0(x^1,x^p),
\Phi(x^1,t)\otimes\Phi(x^p,t)\rangle~
\langle \chi_t^j(x^i)\chi_t^j(x^q) Q_0(x^i,x^q),
\Phi(x^i,t)\otimes\Phi(x^q,t)\rangle .
\nonumber
\eeqn
Each factor here is bounded by
$ C(\Psi)~\De_t/t$.
Similarly to  (\ref{106}),
this can be deduced  from an expression of type (\ref{100rq})
for the factors.
Therefore, the proof of (\ref{112}) reduces to the proof
of the  bound
\be\la{st1}
I_t:=|\langle \chi_t^j(x^1)\ldots\chi_t^j(x^4){\rm
m}^{(4)}_0(x^1,\dots,x^4),
\Phi(x^1,t)\otimes\dots\otimes\Phi(x^4,t)\rangle |\le
C(\Psi) \La^4\De_t^2/t^2,\quad t > 1.
\ee
{\it Step 2}~
Similarly to (\re{er}),
 the estimate (\ref{bphi}) implies,
\be\la{500}
I_t\le
 C~t^{-2d}\sum\limits_{ \bar x }
\chi_t^j(x^1) \ldots \chi_t^j(x^4)
 |{\rm m}^{(4)}_0(x^1,\dots,x^4)|,
\ee
We estimate ${\rm m}^{(4)}_0$
using Lemma  \ref{il} ii).
\begin{lemma}\la{l11.1}
For each $i=2,3,4$ and  all  $\ov x\in W_i$ the
following bound holds:
\be\la{115}
 |{\rm m}^{(4)}_0(x^1,\dots,x^4)|
\le
C \La^4\varphi(|x^1-x^i|/3).
\ee
\end{lemma}
{\bf Proof.}
For $\bar x\in {\cal A}_i$ we apply
 Lemma  \ref{il} ii)
to $\R^{2n}\otimes\R^{2n}$-valued random variables
$\xi=Y_0(x^1)\otimes Y_0(x^p)$ and
 $\eta=Y_0(x^i)\otimes Y_0(x^q)$. Then
(\ref{ass1}) implies the bound  for  all $\bar x\in {\cal A}_i$,
\be\la{M1}
|{\rm m}^{(4)}_0(\bar x)|\le
C \La^4 \varphi(|x^1-x^i|/3).
\ee
For  $\bar x\in {\cal B}_i$,
we apply
 Lemma  \ref{il} ii)
to
$\xi=Y_0(x_1)$ and
 $\eta= Y_0(x^{p})\otimes Y_0(x^{q})\otimes Y_0(x^{i})$. Then
  {\bf S0}  implies a similar bound for
 all $\bar x\in {\cal B}_i$,
\be\la{113}
|{\rm m}^{(4)}_0(\bar x)|=
| M_0^{(4)}(\bar x) -
EY_0(x^1)\otimes
E\Bigl( Y_0(x_{p})\otimes Y_0(x^{q})\otimes Y_0(x^{i})\Bigr)|
\le C \La^4 \varphi(|x^1-x^i|/3),
\ee
and the same for
 all
$\bar x\in {\cal C}_i$. \hfill$\Box$
\medskip\\
{\it Step 3}~It remains to prove the following bounds
for each $i=2,3,4$:
\be\la{vit}
V_i(t):=\sum\limits_{ \bar x }
\chi_t^j(x^1) \ldots \chi_t^j(x^4)
 X_i(\ov x)   \varphi(|x^1-x^i|/3)
\le C\De_t^2 t^{2d-2},
\ee
where $X_i$ is an indicator of the set $W_i$.
In fact, this sum does not depend on $i$,  hence
set $i=2$ in the summand:
\be\la{500s}
V_i(t)
\le
 C\sum\limits_{x^1,x^2 }
\chi_t^j(x^1)\chi_t^j(x^2) \varphi(|x^1-x^2|/3)
\sum\limits_{ x^3 }\chi_t^j(x^3)
\sum\limits_{x^4 }\chi_t^j(x^4)
X_2(\ov x).
\ee
Now a key observation is that the inner sum  in $x^4$
is ${\cal O}(|x^1-x^2|^d)$ as $X_2(\ov x)=0$ for
$|x^4-x^1| > |x^1-x^2|$. This implies
\be\la{500s4}
V_i(t)
\le
 C\sum\limits_{ x^1  }\chi_t^j(x^1)
\sum\limits_{ x^2}\chi_t^j(x^2)
 \varphi(|x^1-x^2|/3)
|x^1-x^2|^d
 \sum\limits_{ x^3 }\chi_t^j(x^3).
\ee
Remember that $\chi^j_t(x)$ is 
an indicator of the room
$R_t^j=\{x\in\Z^d: |x|\le N_t h_t, a^j\le x_d<b^j \}$,
where $N_t=[\ga t/h_t]$.
The inner sum in $x^2$ is bounded as
\beqn\la{qphi4}
&&\int\limits_{ |x^2|\le \ga t}
 \varphi(|x^1-x^2|/3)
|x^1-x^2|^d \, dx^2
\le C(d)\int\limits_{0}^{2\ga t}r^{2d-1} \varphi( r/3)\,dr
\nonumber\\
&&\le
C_1(d) \sup\limits_{r\in [0,2\ga t]}
 ~r^d\varphi^{1/2}( r/3)
\int\limits_{0}^{2\ga t}r^{d-1} \varphi^{1/2}(  r/3)\,dr,
\eeqn
where the `$\sup$' and the last integral
 are bounded by
(\re{1111}) and (\re{1.12}), respectively. Therefore,  (\re{vit})
follows from (\re{500s4}).
This completes the proof of
Theorem \ref{p5.1}.
\hfill$\Box$



\setcounter{equation}{0}
\section{Appendix. Dynamics and covariance
in  Fourier space }

{\bf Proof of Proposition \ref{p1.1}}
Applying Fourier transform
to (\ref{CP})
we obtain
\be\la{CPF}
\dot{\hat Y}(t)=
\hat {\cal A}(\theta)\hat Y(t),\,\,\,t\in\R,
\,\,\,\,\hat Y(0)=\hat Y_0.
\ee
Here we  denote
\be\la{hA}
\hat{\cal A}(\theta)=\left(
 \begin{array}{cc}
0 & 1\\
-\hat V(\theta) & 0
\end{array}\right),\,\,\,\,\theta\in T^d.
\ee
Note that $\hat Y(\cdot,t)\in D'(T^d)$
 for $t\in\R$.
On the other hand,
 $\hat V(\theta)$
is a smooth function by {\bf E1}.
Therefore,
the solution  $\hat Y(\theta,t)$ of (\ref{CPF})
exists,  is unique and admits the representation
$\hat Y(\theta,t)=
\exp\Big(\hat{\cal A}(\theta)t\Big)\hat Y_0(\theta)$
which becomes
the convolution
\be\la{solGr}
Y(x,t)=\sum\limits_{x'\in\Z^d}{\cal G}_t(x-x')
Y_0(x')
\ee
 in the coordinate space, where the
  Green function ${\cal G}_t(z)$ admits the
Fourier representation
\be\la{Grcs}
{\cal G}_t(z):=
F^{-1}_{\theta\to z}[
\exp\big(\hat{\cal A}(\theta)t\big)]
=(2\pi)^{-d}\int\limits_{T^d}e^{-iz\theta}
\exp\big(\hat{\cal A}(\theta)t\big)d\theta.
\ee
Hence, by the partial integration,
 ${\cal G}_t(z)\sim|z|^{-p}$
as $|z|\to\infty$ for any
$p >0$ and bounded $|t|$
since $\hat{\cal A}(\theta)$
is the smooth function of $\theta\in T^d$.
Therefore,
the convolution representation (\ref{solGr})
implies $Y(t)\in {\cal H}_\al$.
\hfill$\Box$\\
~\\
{\bf  Covariance in  Fourier space}
Note that $\hat{\cal G}_t( \theta)$
has a form
\be\la{hatcalG}
\hat{\cal G}_t( \theta)=
\left( \begin{array}{cc}
 \cos\Om t &~ \sin \Om t~\Om^{-1}  \\
 -\sin\Om t~\Om
&  \cos\Om t\end{array}\right),
\ee
where $\Om=\Om( \theta)$ is the
Hermitian matrix defined by (\ref{Omega}).
Let $\hat C(\theta)$ be defined by (\ref{C(theta)})
and  $I$ be the identity matrix.
Then
\be\la{Gtdec}
\hat{\cal G}_t( \theta)=\cos\Om t\, I+
\sin\Om t\, \hat C(\theta).
\ee
Denote by $Q(x,y):=E\Big(Y_0(x)\otimes Y_0(y)\Big)$,
and $Q_t(x,y):=E\Big(Y(x,t)\otimes Y(y,t)\Big)$.
Hence,
applying Fourier transform
to $Q_t(x,y)$ we get
$$
\hat Q_t(\theta,\theta'):=
F_{x\to\theta, y\to-\theta'}Q_t(x,y)=
\hat{\cal G}_t(\theta)\hat Q(\theta,\theta')\hat{\cal G}^T_t(-\theta'),
$$
where
$\hat Q(\theta,\theta'):=
F_{x\to\theta, y\to-\theta'}Q(x,y)$.
Note that
due to condition {\bf E2}
 $\Om^T(-\theta')=
\Om^*(\theta')=\Om(\theta')$
and then
$\hat{\cal G}^T_t(-\theta')=
\hat{\cal G}^*_t(\theta')$,
where
\be\la{G*tdec}
\hat{\cal G}^*_t( \theta):=\cos\Om t\, I+
\sin\Om t\, \hat C^*(\theta).
\ee
Here
$\hat C^*$ is Hermitian
adjoint matrix to $\hat C$
 as in (\ref{C(theta)}).
Then
\beqn\la{ap1}
\hat Q_t(\theta,\theta')
&=&
\cos\Om(\theta) t~\hat Q(\theta,\theta')\, \cos\Om(\theta')t+
\sin\Om(\theta)t~\hat C(\theta)\hat Q(\theta,\theta')
\hat C^*(\theta')\, \sin\Om(\theta') t
\nonumber\\
&&+\cos\Om(\theta) t~\hat Q(\theta,\theta')
\hat C^*(\theta')\, \sin\Om(\theta') t
+
\sin\Om(\theta) t~\hat C(\theta)
\hat Q(\theta,\theta')\,\cos\Om(\theta')t.\,\,\,\,\,\,\,\,\,\,\,
\eeqn
Now, for simplicity of calculations, we 
will assume that
the set of the `crossing' points $\theta_*$
is empty, i.e. $\om_k(\theta)\ne\om_l(\theta)$,
$\forall k,l\in\ov n$, and
the functions 
$\om_k(\theta)$ and $B(\theta)$ are real-analytic.
For example, 
this is the case of the  simple elastic lattice
(\ref{dKG}).
(Otherwise, we need a partition of unity
(\ref{part})). 
Consider the first term in the RHS of (\ref{ap1}).
We rewrite it using
 (\ref{diag}) 
in the form
\beqn
&&\cos\Om(\theta) t~\hat Q(\theta,\theta')\, \cos\Om(\theta')t
=B(\theta)\Big(\cos\om_k(\theta)t~
A(\theta,\theta')_{kl}\cos\om_l(\theta')t
\Big)_{k,l\in \bar n}B^*(\theta')
\nonumber\\
~&&\nonumber\\
&&=
B(\theta) \frac{1}{2}\Big( (\cos(\om_k(\theta)\!-\!\om_l(\theta'))t +
\cos(\om_k(\theta)\!+\!\om_l(\theta'))t) A(\theta,\theta')_{kl}
\Big)_{k,l\in \bar n}B^*(\theta'),\,
\eeqn
where
$A(\theta,\theta')
:=B^*(\theta)\hat Q(\theta,\theta')
B(\theta')$.
Similarly, we can rewrite
the remaining three terms in the RHS of (\ref{ap1}).
Finally,
\beqn\la{ap.11}
\hat Q_t(\theta,\theta')
=
\hat{\cal G}_t(\theta)\hat Q(\theta,\theta')\hat{\cal G}^*_t(\theta')
=
B(\theta)R_t(\theta,\theta') B^*(\theta'),
\eeqn
where by $R_t(\theta,\theta')$
we denote the $2n\times 2n$ matrix
with the entries
\beqn\la{Rkl}
R_t(\theta,\theta')_{kl}&:=&\frac{1}{2}
 \sum\limits_{\pm}
\Big\{
\cos\big(
\om_k(\theta)\!\pm\!\om_l(\theta')\big)t
~\Big[
B^*(\theta)
\Big(
\hat Q(\theta,\theta')\!\mp\!\hat C(\theta)\hat Q(\theta,\theta')\hat
C^*(\theta')\Big)B(\theta')\Big]_{kl}
\nonumber\\
&&+
\sin\big(\om_k(\theta)\!\pm\!\om_l(\theta')\big)t
~\Big[
B^*(\theta)
\Big(
\hat C(\theta)\hat Q(\theta,\theta')\!\pm\!\hat Q(\theta,\theta')\hat
C^*(\theta')
\Big)B(\theta')\Big]_{kl}\Big\}.\,\,\,\,\,\,\,\,\,\,\,
\eeqn
In the translation-invariant  case $Q(x,y)=q(x-y)$,
$\hat Q(\theta,\theta')=\delta(\theta-\theta')\hat q(\theta)$
and  we get
\beqn\la{ap22}
\hat Q_t(\theta,\theta')=
\delta(\theta-\theta')
B(\theta)R_t(\theta) B^*(\theta),
\eeqn
where
 by $R_t(\theta)$
we denote the $2n\times 2n$ matrix
with the entries
\beqn\la{ap.2}
R_t(\theta)_{kl}&=&\frac{1}{2}\sum\limits_{\pm}\Big\{
\cos\big(
\om_k(\theta)\!\pm\!\om_l(\theta)\big)t
~\Big[
B^*(\theta)
\Big(
\hat q(\theta)\mp\hat C(\theta)\hat q(\theta)\hat
C^*(\theta)\Big)B(\theta)\Big]_{kl}
\nonumber\\
&&+
\sin\big(\om_k(\theta)\!\pm\!\om_l(\theta)\big)t
~\Big[
B^*(\theta)
\Big(
\hat C(\theta)\hat q(\theta)\pm\hat q(\theta)\hat
C^*(\theta)
\Big)B(\theta)\Big]_{kl}\Big\}.
\eeqn 
Let us denote by $p(\theta):=B^*(\theta)\hat q(\theta)B(\theta) $.
Then by (\ref{diag}) and (\ref{C(theta)})
we obtain
\beqn\la{ap.4} 
R_t(\theta)_{kl}&=& 
\frac{1}{2}\sum\limits_\pm
\Big\{\cos\big( 
\om_k(\theta)\!\mp\!\om_l(\theta)\big)t 
~\left(
\ba{cc}
p^{00}_{kl}\pm\om_k^{-1}p^{11}_{kl}\om_l^{-1}&
p^{01}_{kl}\mp\om_k^{-1}p^{10}_{kl}\om_l\\
p^{10}_{kl}\mp\om_k p^{01}_{kl}\om_l^{-1}&
p^{11}_{kl}\pm\om_k p^{00}_{kl}\om_l
\ea
\right)
\nonumber\\ 
&&+ 
\sin\big(\om_k(\theta)\!\pm\!\om_l(\theta)\big)t 
~\left(
\ba{cc}
\om_k^{-1}p^{10}_{kl}\pm p^{01}_{kl}\om_l^{-1}&
\om_k^{-1}p^{11}_{kl}\mp p^{00}_{kl}\om_l\\
-\om_kp^{00}_{kl}\pm p^{11}_{kl}\om_l^{-1}&
-\om_kp^{01}_{kl}\mp p^{10}_{kl}\om_l
\ea
\right)\Big\}.
\eeqn 
We enumerate the eigenvalues $\om_k(\theta)$ 
as in (\ref{enum}). Then 
$\cos\big(\om_k(\theta)-\om_l(\theta)\big)t=1$ 
 for  $k,l\in(r_{\sigma-1},r_\sigma]$,
 $\sigma=1,\dots,s+1$, hence 
\beqn\la{ap.3} 
R_t(\theta)_{kl}&=& 
\frac{1}{2}\Big[B^*(\theta) 
\Big( 
\hat q(\theta)+\hat C(\theta)\hat q(\theta)\hat 
C^*(\theta)\Big)B(\theta)\Big]_{kl} 
\nonumber\\ 
&&+ 
\frac{1}{2}\cos 2\om_k(\theta)t 
~\Big[ 
B^*(\theta) 
\Big( 
\hat q(\theta)-\hat C(\theta)\hat q(\theta)\hat 
C^*(\theta) 
\Big)B(\theta)\Big]_{kl} 
\nonumber\\ 
&&+ 
\frac{1}{2}\sin2\om_k(\theta)t 
~\Big[ 
B^*(\theta) 
\Big( 
\hat C(\theta)\hat q(\theta)\!+\! 
\hat q(\theta)\hat C^*(\theta) 
\Big)B(\theta)\Big]_{kl}. 
\eeqn

 
\end{document}